\documentclass[letterpaper]{aa}
\usepackage{txfonts}
\usepackage{graphicx}
\usepackage{t1enc}

\newcommand{\Td}    {T_\mathrm{d}}
\newcommand{\Tex}   {T_\mathrm{ex}}
\newcommand{\Trot}   {T_\mathrm{rot}}

\newcommand{\kms}   {km~s$^{-1}$}
\newcommand{\cmt}   {cm$^{-3}$}
\newcommand{\jpb}   {$\rm Jy~beam^{-1}$}	
\newcommand{\lo}    {$L_{\sun}$}
\newcommand{\mo}    {$M_{\sun}$}
\newcommand{\nh}    {NH$_3$}
\newcommand{\nth}   {N$_2$H$^+$}
\newcommand{\chtoh} {CH$_3$OH}

\newcommand{\et}    {et al.}
\newcommand{\eg}    {e.\,g.,}
\newcommand{\uchii}    {UC\ion{H}{ii}}

\begin{document}

\title{Star formation in a clustered environment \\
around the \uchii\ region in IRAS~20293+3952}

\author{
A. Palau$^1$, R. Estalella$^1$, J. M. Girart$^{2}$, P. T. P. Ho$^3$, 
Q. Zhang$^3$, \& H. Beuther$^4$
}

\offprints{Aina Palau,\\ \email{apalau@am.ub.es}}

\institute{
Departament d'Astronomia i Meteorologia, Universitat de Barcelona,
Av.\ Diagonal 647, 08028 Barcelona, Catalunya, Spain
\and
Institut de Ci\`encies de l'Espai (CSIC-IEEC), Campus UAB, Facultat de
Ci\`encies, Torre C5-Parell-2a, 08193 Bellaterra, Catalunya, Spain
\and
Harvard-Smithsonian Center for Astrophysics,
60 Garden Street, Cambridge, MA 02138, USA
\and
Max-Planck-Institut for Astronomy, Koenigstuhl 17, 69117 Heidelberg, Germany
}

\date{Received / Accepted}

\authorrunning{Palau \et}
\titlerunning{
Star Formation in IRAS 20293+3952
}

\abstract{}{We aim at studying the cluster environment surrounding the \uchii\
region in IRAS~20293+3952, a region in the first stages of formation of a
cluster around a high-mass star.}{BIMA and VLA were used to observe the 3~mm
continuum, \nth\,(1--0), \nh\,(1,1), \nh\,(2,2), and \chtoh\,(2--1) emission of
the surroundings of the \uchii\ region. We studied the kinematics of the region
and computed the rotational temperature and column density maps by fitting the
hyperfine structure of \nth\ and \nh.}{The dense gas traced by N$_2$H$^+$ and
NH$_3$ shows two different clouds, a main cloud to the east of the \uchii\
region, of $\sim$0.5~pc and $\sim$250~\mo, and a western cloud, of
$\sim$0.15~pc and $\sim$30~\mo. The dust emission reveals two strong components
in the northern side of the main cloud,  BIMA~1 and BIMA~2, associated with
YSOs driving molecular outflows, and two fainter components in the southern
side, BIMA~3 and BIMA~4, with no signs of star forming activity. Regarding the
CH$_3$OH, we found strong emission in a fork-like structure associated with
outflow B, as well as emission associated with outflow A. The YSOs associated
with the dense gas seem to have a diversity of age and properties. The
rotational temperature is higher in the northern side of the main cloud, around
22~K, where there are most of the YSOs, than in the southern side, around 16
K.   There is strong chemical differentiation in the region, since we
determined low values of the NH$_3$/N$_2$H$^+$ ratio, $\sim$50, associated with
YSOs in the north of the main cloud, and high values, up to 300, associated
with cores with no detected YSOs, in the south of the main cloud. Such a
chemical differentiation is likely due to abundance/depletion effects. Finally,
interaction between the different sources in the region is important. First,
the \uchii\ region is interacting with the main cloud, heating it and enhancing the
CN\,(1--0) emission. Second, outflow A seems to be excavating a cavity and
heating its walls. Third, outflow B is interacting with the BIMA~4 core, likely
producing the deflection of the outflow and illuminating a clump located
$\sim$0.2~pc to the northeast of the shock.}{The star formation process in
IRAS~20293+3952 is not obviously associated with interactions, but seems to
take place where density is highest.}


\maketitle

\section{Introduction}

It is generally accepted that the formation of massive stars takes place not
isolated but simultaneously with the formation of a cluster.  Clusters of
infrared sources have been studied at optical and near-infrared wavelengths
around intermediate/high-mass stars (e.g., McCaughrean \& Stauffer 1994;
Hillenbrand 1995;  Hillenbrand \& Hartmann 1998; Testi et al.\ 1998, 1999,
2000). One could surmise on the structure and the evolutionary state of the
molecular cloud when such clusters were being born around an
intermediate/high-mass protostar. In the standard theory of star formation, the
central massive object, deeply embedded in large amounts of gas and dust,
starts to radiate at UV wavelengths at very early stages, before finishing the
accretion of all its final mass (\eg\ Bernasconi \& Maeder 1996). The UV
photons ionize the surrounding medium, and the high-mass star developes an
\uchii\ region around it (see, eg. Garay \& Lizano 1999), which expands and
pushes forward the gas and dust surrounding it until the parental cloud is
completely disrupted and a cluster of infrared/optical sources emerges (\eg\
Hester \& Desch 2005). The study of the clumpy medium around massive protostars
helps not only in the understanding of the formation of the massive star
itself, but also provides a direct way to study the clustered mode of star
formation, in which most stars are thought to form.

In order to properly characterize the medium around a massive protostar, it is
necessary to be sensitive to small ($\sim$2000~AU) and  low-mass ($\la 1$~\mo)
clumps of gas and dust. For this, interferometric observations toward
relatively nearby ($<3$~kpc) massive star-forming regions are required, and are
now possible with sufficient sensitivity.  Here we report on observations of
one intermediate/high-mass star-forming region, IRAS~20293+3952, located at
2.0~kpc of distance (Beuther \et\ 2004b) in the southwestern edge of the
Cygnus~OB2 association, and with a luminosity of 6300~\lo. This region has been
included in the sample of  high-mass protostar candidates of Sridharan et al.\
(2002). The IRAS source is associated with the only centimeter source detected
in the region, which is likely tracing an \uchii\ region (Beuther et al.\
2002b). A study of the H$_2$ emission in the region shows two near-infrared
stars (IRS~1 and IRS~2) associated with the IRAS source, and a circular ring of
H$_2$ emission surrounding IRS~1 (Kumar \et\ 2002). Beuther et al.\ (2002a)
observed the region with the IRAM~30\,m telescope and find some substructure at
1.2~mm around the \uchii\ region. The strongest millimeter peak, very close to
the position of a H$_2$O maser, is located $\sim$15$''$  north-east of the IRAS
source, and observations with high angular resolution by Beuther \et\ (2004a)
reveal a compact and strong millimeter source, mm1, associated with the H$_2$O
maser. Two other fainter compact millimeter sources, mm2 and mm3, are located
$10''$ to the east of the \uchii\ region. Beuther \et\ (2004a), from 
CO\,(2--1) and SiO\,(2--1) observations, suggest the presence of four molecular
outflows in the region, with two of them, outflows A and B, associated with two
chains of H$_2$ knots (Kumar \et\ 2002). Subsequent observations with the PdBI
were carried out at 2.6 and 1.3 mm by  Beuther \et\ (2004b), who find CN
emission close to mm1, mm2 and mm3.  The different millimeter sources detected
and the presence of multiple outflows around the \uchii\ region makes this
region a good choice to study star formation and interaction in a clustered
environment.



In this paper we report on BIMA and VLA observations of the continuum at 3~mm
and the dense gas traced by \nth\,(1--0), \nh\,(1,1), and \nh\,(2,2) toward
IRAS~20293+3952, together with observations of several \chtoh\,(2--1)
transitions.  In \S\,2 we summarize the observations and the reduction process,
in \S\,3 we show the main results for the continuum and molecular line
emission, in \S\,4 we analyze the line emission and show the method used
to derive the rotational temperature and column density maps. Finally, in \S\,5
we discuss the results obtained, mainly the properties of the dense gas
surrounding the \uchii\ region, the different sources identified in the region,
and the interaction between them.




\begin{table*}{\footnotesize
\caption{Parameters of the BIMA and VLA observations \label{tparobs}}
\centering
\begin{tabular}{lcccccc}
\hline\hline
&&
&Beam
&P.A.
&Spec. resol.
&Rms$^\mathrm{a}$
\\
Observation 
&Telescope
&Config.
&(arcsec)
&(\degr)
&(\kms)
&(m\jpb)\\
\hline
Continuum
&BIMA
&B+C
&$5.8 \times 5.6$
&$-6.4$
&$-$
&0.7
\\
\chtoh\,(2--1)
&BIMA
&B+C
&$6.1 \times 5.7$
&$-6.2$
&0.31
&80
\\
\nth\,(1--0)
&BIMA
&B+C
&$6.2 \times 5.9$
&$-4.1$
&0.31
&70
\\
\nh\,(1,1)
&VLA
&D
&$6.9 \times 3.0$
&71.5
&0.62
&3
\\
\nh\,(2,2)
&VLA
&D
&$6.6 \times 3.1$
&71.4
&0.62
&3
\\
\hline
\end{tabular}
\begin{list}{}{}
\item[$^\mathrm{a}$] Rms noise per channel in the case of line emission.
\end{list}
}
\end{table*}

\section{Observations}

\subsection{BIMA}

The \nth\,(1--0) and \chtoh\,(2--1) lines and continuum at 95~GHz were observed
towards the IRAS~20293+3952 region with the BIMA array\footnote{The BIMA array
was operated by the Berkeley-Illinois-Maryland Association with support from
the National Science Foundation.} at Hat Creek. The observations were carried
out on 2003 September 28 in the C configuration, and on 2004 March 24, in the B
configuration, with 10 antennas in use.  The phase center of the observations
was set at $\alpha(J2000)=20^{\mathrm h}31^{\mathrm m} 12\fs70$;
$\delta(J2000)=+40\degr03\arcmin13\farcs4$ (position of the millimeter peak
detected with the IRAM~30\,m telescope by Beuther \et \ 2002a). The full range
of projected baselines, including both configurations, was 9.4--240~m. The FWHM
of the primary beam at the frequency of the observations was $\sim$120$''$.
Typical system temperatures were $\sim$500~K for Sep 28 and 500--1500 K for Mar
24.

The digital correlator was configured to observe simultaneously the continuum
emission, the \nth\,(1--0) group of hyperfine transitions (93.17631 GHz, in
the lower sideband), and the \chtoh\ 2$_1$--1$_1$ E, 2$_0$--1$_0$ E,
2$_0$--1$_0$ A$^+$, and 2$_{-1}$--1$_{-1}$ E transitions (96.74142 GHz, in the
upper sideband). For the continuum, we used a bandwidth of 600 MHz in each
sideband, and for the lines we used a bandwidth of 25 MHz with 256 channels of
100~kHz width, providing a spectral resolution of 0.31~\kms. 

Phase calibration was performed with QSO~2013+370, with typical rms in the
phases of 12\degr\ and 43\degr\ for C and B configurations, respectively. The
absolute position accuracy was estimated to be around $0\farcs5$. We also used 
QSO~2013+370 for flux calibration, and the error in the flux density scale was
assumed to be $\sim$30\%. Data were calibrated and imaged using standard
procedures in MIRIAD (Sault \et\ 1995). We combined the data from B and C
configurations. In order to improve the angular resolution of the continuum
emission, we weighted the visibility data with a robust parameter of $+1$.  For
the line emission, which is more extended than the continuum, we used natural
weighting.  The resulting synthesized beams  and the final rms noises are
listed in Table~\ref{tparobs}.



\subsection{VLA}

Observations of the $(J,K)=(1,1)$ and $(2,2)$ inversion transitions of the
ammonia molecule were carried out with the Very Large Array  (VLA) of the
NRAO\footnote{The National Radio Astronomy Observatory is a facility of the
National Science Foundation operated under cooperative agreement by Associated
Universities, Inc.} in the D configuration on 2000 September 3. The phase
center was set to $\alpha(J2000)=20^{\mathrm h}31^{\mathrm m}10\fs70$;
$\delta(J2000)=+40\degr03\arcmin10\farcs7$ (catalogue position of the IRAS
source). The FWHM of the primary beam at the frequency of observation was
$\sim$110$''$, and the range of projected baselines was 37.9--708~m. The
absolute flux calibration was performed by using 3C\,48, adopting a flux
density at 1.3 cm of 1.05 Jy. The phase calibrator was QSO~2013+370, with a
1.3~cm bootstrapped flux density of 3.84~Jy, and 3C\,84 was used as the
bandpass calibrator.


The \nh\,(1,1) and \nh\,(2,2) lines were observed simultaneously in the 4 IF
correlator mode of the VLA (with 2 polarizations for each line), providing 63
channels with a spectral resolution of 0.62~\kms \ across a bandwidth of
3.13~MHz, plus an additional continuum channel containing the central 75\% of
the total bandwidth. The bandwidth was centered at the systemic velocity
$v_\mathrm{LSR}=6.3$~\kms\ (Sridharan \et\ 2002) for the \nh\,(1,1) line, and
at $v_\mathrm{LSR}=11.3$ \kms \ for the \nh\,(2,2) line (to cover the main and
one of the satellite components). Data were calibrated and imaged using
standard procedures of AIPS. The cleaned channel maps were made using natural
weighting of the visibility data. Table~\ref{tparobs} summarizes the parameters
of the observations.


\begin{figure*}
\begin{center}
\includegraphics[scale=0.8]{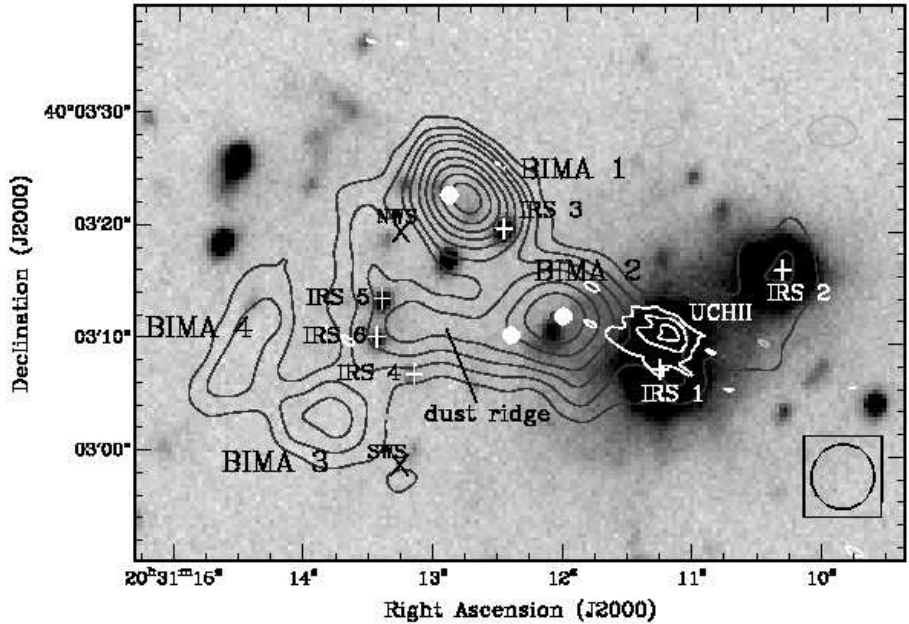}
\caption{
Contours: 3.15 mm continuum emission from IRAS~20293+3952 obtained with BIMA in
the B and C configurations using an intermediate weight between uniform and
natural ($\mathrm{robust}=+1$). The contour levels are  $-3$, 3, 5, 7, 9, 12,
15, 18, 21, and 24 times the rms noise of the map,  0.67~mJy~beam$^{-1}$. The
synthesized beam, shown in the bottom right corner, is $5\farcs8 \times
5\farcs6$, at $\mathrm{P.A.}=-6\fdg4$.  White contours correspond to the
centimeter emission at 3.6 cm, and trace the \uchii\ region (Beuther \et\ 2004a).
Contours are $-3$, 3, 6, and 9 times 70~$\mu$Jy~beam$^{-1}$. 
Grey scale: Continuum image at 2.12 $\mu$m from Kumar \et \ (2002).  Main
clumps of dust emission are labeled as BIMA~1 to BIMA~4. White dots indicate
the compact millimeter sources detected by Beuther \et\ (2004b) at 2.6 mm with
the PdBI. These are, from west to east, mm2, mm3, and mm1. White crosses mark
the sources from 2MASS with infrared excess, and black tilted crosses mark the
position of the Northern Warm Spot (NWS) and the Southern Warm Spot (SWS).
\label{fcontr}
}
\end{center}
\end{figure*}

\section{Results}

\subsection{Continuum emission \label{srcont}}


Figure \ref{fcontr} shows the continuum map at 3.15 mm overlaid on the
continuum infrared emission at 2.12 $\mu$m from Kumar \et\ (2002).  The
millimeter continuum emission has two strong components, BIMA~1 and BIMA~2,
both spatially resolved. While BIMA~1 is elongated in a northeast-southwest
direction (roughly P.A.$\simeq$45\degr), BIMA~2 is somewhat flattened in the
east-west direction, and is surrounded by an elongated structure of
$\sim$30$''$ (0.3~pc), which is tracing a dust ridge.  There is an extended
clump $10''$ to the southeast of the dust ridge, which we label BIMA~3. Note
that in this southeastern region there are very few infrared sources.  At
$10''$ to the east of the ridge, there is another faint feature, BIMA~4,
extending for $10''$ in the northwest-southeast direction.  Continuum emission
at a level of 5$\sigma$ is detected toward the position of the \uchii\ region,
and toward IRS~2.    Beuther \et\ (2004b) observed the same region at 2.6 mm
with the PdBI at higher angular resolution ($1\farcs5 \times 1\farcs2$), but 
with a smaller primary beam ($44''$). The authors find a very strong compact
source, mm1, toward the position of BIMA~1 and elongated in the same
northeast-southwest direction as BIMA~1. Besides this strong component, a weak
subcomponent was resolved in the direction of the elongation (to the southwest
of mm1), and was labeled mm1a. Beuther \et\ (2004b) also find two compact
millimeter sources associated with BIMA~2, mm2 and mm3 (see Fig.~\ref{fcontr}
for the positions of these compact millimeter sources).

\begin{table*}{\footnotesize
\caption{Parameters of the continuum sources at 3.15~mm \label{tcont}}
\centering
\begin{tabular}{lcccccc}
\hline\hline
&\multicolumn{2}{c}{Peak Position}
&$I_\nu^\mathrm{peak}$
&Flux$^\mathrm{a}$
&Assumed $\Td$$^\mathrm{b}$
&Mass$^\mathrm{c}$
\\
&$\alpha (\rm J2000)$
&$\delta (\rm J2000)$
&(m\jpb)
&(mJy)
&(K)
&(\mo)
\\
\hline
BIMA~1
&20:31:12.767
&40:03:22.65
&17.4
&28
&34
&6.3
\\
BIMA~2
&20:31:12.062
&40:03:12.22
&12.1
&23
&34
&5.2
\\
BIMA~2+ridge
&-
&-
&-
&42
&25
&14
\\
BIMA~3
&20:31:13.769
&40:03:03.12
&5.4
&7.9
&17
&3.9
\\
BIMA~4
&20:31:14.558
&40:03:06.95
&4.2
&7.9
&17
&3.9
\\
IRS~2
&20:31:10.303
&40:03:17.10
&3.5
&3.9
&50
&0.60
\\
\hline
\end{tabular}
\begin{list}{}{}
\item[$^\mathrm{a}$] Flux density inside the 5$\sigma$ contour level for
BIMA~1, BIMA~2, and BIMA~2+ridge, and the 3$\sigma$ level for BIMA~3, 
BIMA~4, and IRS~2.
\item[$^\mathrm{b}$] $\Td$ is estimated by correcting the rotational
temperature derived from \nh\ (see \S~\ref{satrot}) to kinetic temperature,
following the expression of Tafalla \et\  (2004). For IRS~2 we assumed $\Td \sim 50$~K.
\item[$^\mathrm{c}$] Masses of gas and dust derived assuming a dust emissivity
index $\beta=1$ (Beuther \et\ 2004b). The uncertainty in the masses due to the opacity
law and the dust emissivity index is estimated to be a factor of four.
\end{list}
}
\end{table*}

In Table \ref{tcont} we show the position, peak intensity, flux density and
masses associated with BIMA~1, BIMA~2, the entire dust ridge (including BIMA
2), BIMA~3, BIMA~4, and IRS~2. We discuss these results in section \ref{sad}.

\begin{figure*}
\begin{center}
\includegraphics[scale=0.7]{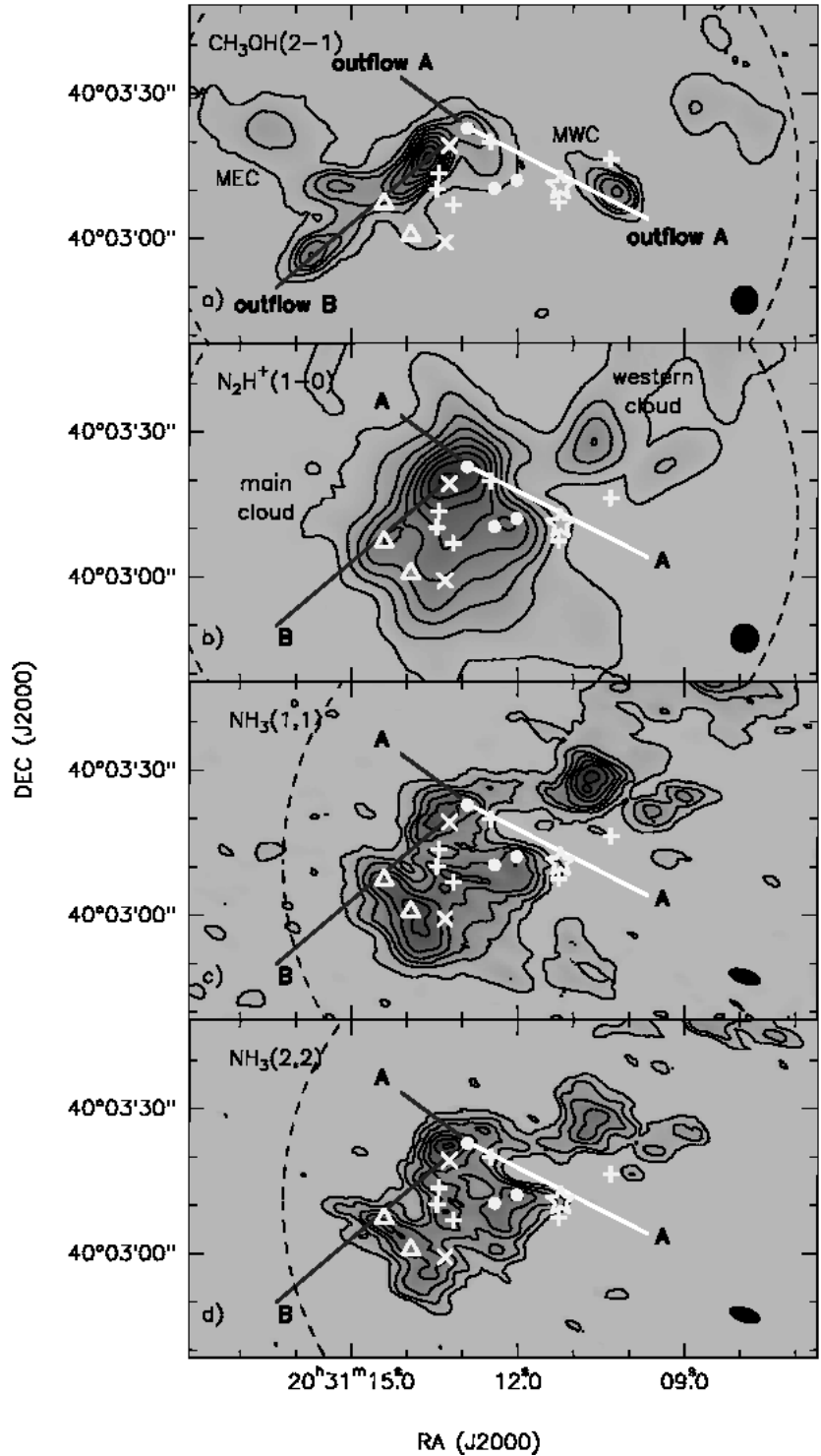}
\end{center}
\caption{
{\bf a)} \chtoh\ zero-order moment for the 2$_0$--1$_0$ A$^+$, and
2$_{-1}$--1$_{-1}$ E lines toward IRAS~20293+3952. Contours start at 1\%, 
increasing in steps of 12\% of the peak intensity, 10.7~Jy~beam$^{-1}$~\kms.  
{\bf b)} \nth\ zero-order moment integrated for all the hyperfine components
of the (1--0) transition. Contours start at 2\%, increasing in steps of 12\% of
the peak intensity, 21.9~Jy~beam$^{-1}$~\kms.  
{\bf c)} \nh\,(1,1) zero-order moment. Contours start at 5\%, increasing in
steps of 15\% of the peak intensity,  0.240~Jy~beam$^{-1}$~\kms.  
{\bf d)} \nh\,(2,2) zero-order moment. Contours start at 7\%, increasing in
steps of 15\% of the peak intensity, 0.0892~Jy~beam$^{-1}$~\kms.   In all
panels, the first level is about 3 times the rms noise of the map. Symbols are
the same as in Fig.~\ref{fcontr}, with the star marking the peak of the
centimeter emission, and the triangles marking the positions of BIMA~3 and
BIMA~4 cores. MEC and MWC stand for methanol eastern clump and methanol western
clump, respectively. The synthesized beams for each transition are shown in the
bottom right corner, and are listed in Table \ref{tparobs}. The straight lines
mark the direction of outflows A and B  (Beuther \et\ 2004a), with the
light grey line corresponding to the redshifted lobe. The dashed curve
indicates the primary beam of BIMA (panels a and b) and VLA (panels c and d)
observations.
\label{fm0}
}
\end{figure*}

\subsection{CH$_3$OH \label{srch3oh}}

\begin{figure*}
\begin{center}
\includegraphics[scale=0.6,angle=0]{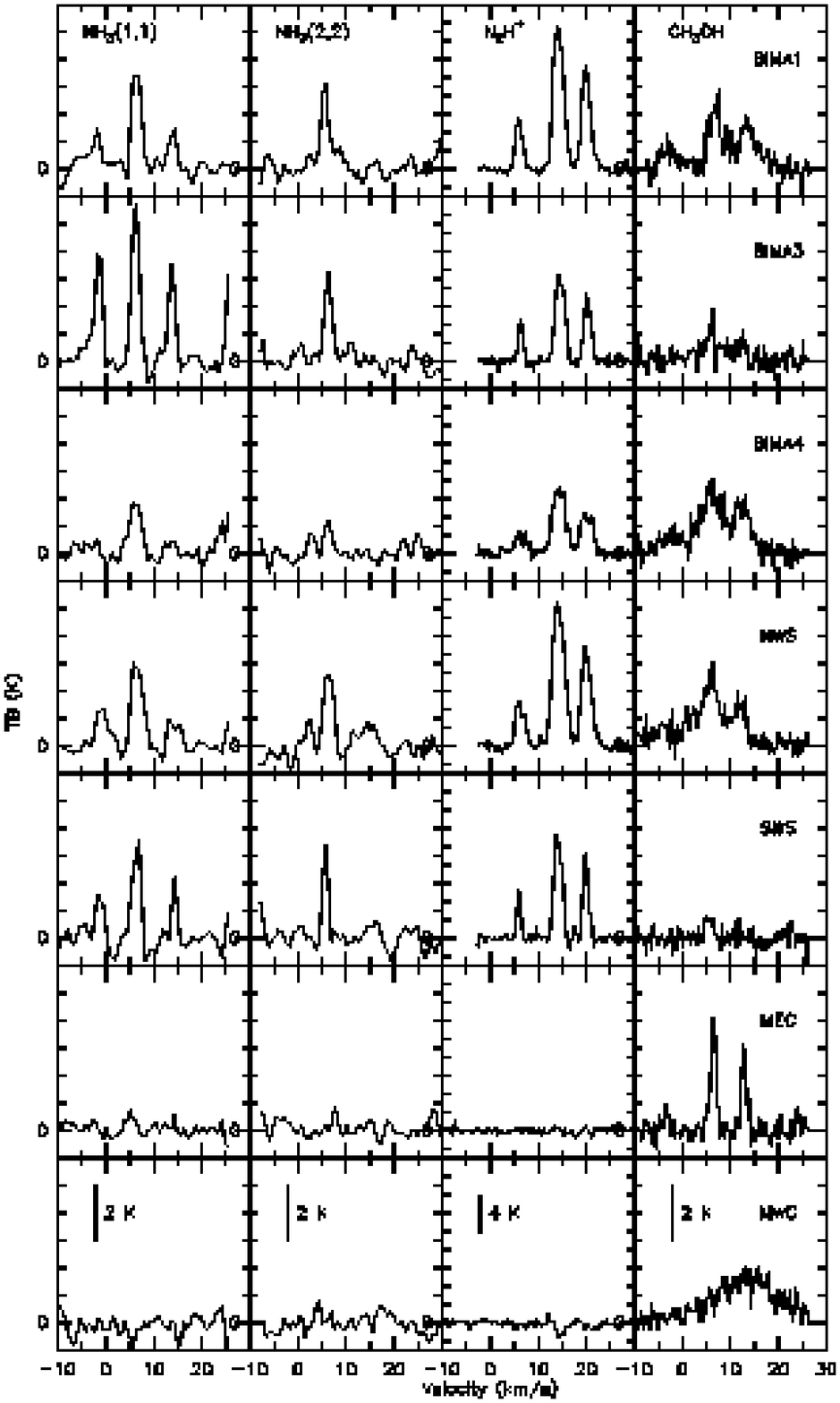}
\caption{
Spectra toward seven positions of the IRAS~20293+3952 region for the four
transitions studied in this paper, from left to right, \nh\,(1,1), \nh\,(2,2),
\nth\,(1--0), and \chtoh\,(2--1). The seven positions are labeled on the right
panel of each row,  and are, from top to bottom, BIMA~1, BIMA~3, BIMA~4, NWS
(Northern Warm Spot), SWS (Southern Warm Spot), MEC (methanol eastern clump), and
MWC (methanol western clump). The vertical scale is indicated for each
transition in the bottom row. For \chtoh\ we show, in order of increasing
velocity, the 2$_0$--1$_0$ E, 2$_0$--1$_0$ A$^+$, and 2$_{-1}$--1$_{-1}$ E
lines (velocities are referred to the 2$_0$--1$_0$ A$^+$ line). For \nth,
velocities are referred to the $F_1=0$--1 hyperfine.
\label{fspecs}
}
\end{center}
\end{figure*}

\begin{figure*}
\begin{center}
\includegraphics[scale=0.9]{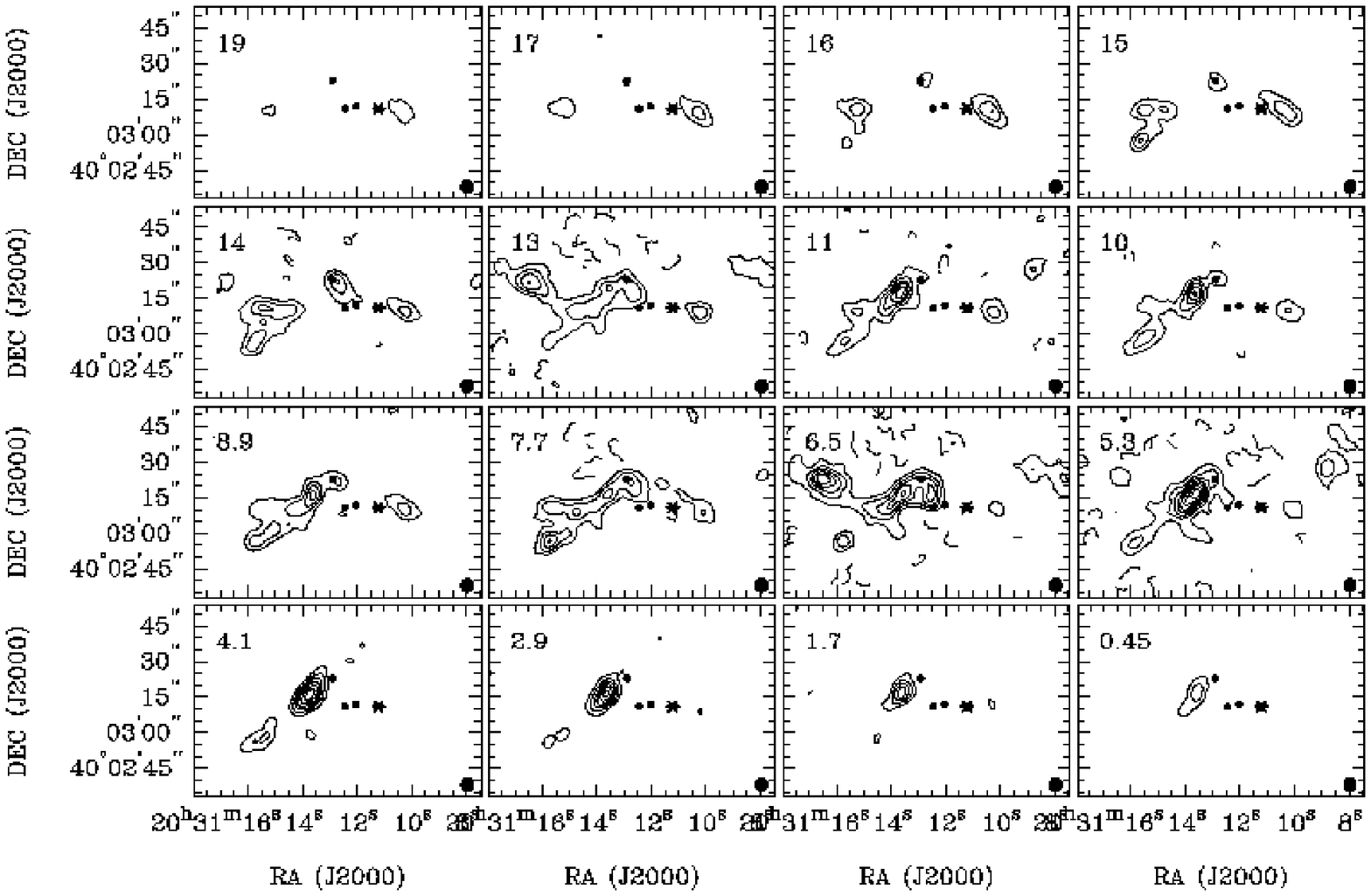}
\caption{
\chtoh\ channel maps toward IRAS~20293+3952 for the lines 2$_0$--1$_0$ A$^+$
(around 6.3 \kms, the systemic velocity), and 2$_{-1}$--1$_{-1}$ E (around 13
\kms), averaged every 4 channels. Contours are $-16$, $-12$, $-8$, $-4$, 4, 8,
12, 16, 20, 24, 28, and 32 times the rms of the map, 0.04~Jy~beam$^{-1}$.
Filled circles are the compact millimeter sources detected by Beuther \et\
(2004b), and the star marks the position of the \uchii\ region. The synthesized
beam is shown in the bottom right corner of each panel.
\label{fchch3oh}
}
\end{center}
\end{figure*}

Figure~\ref{fm0}a shows the zero-order moment map of the \chtoh \ emission,
integrated from $-1.5$ to 21 \kms, including the two strongest transitions
2$_0$--1$_0$ A$^+$, and 2$_{-1}$--1$_{-1}$ E.  Figure~\ref{fspecs}  shows the
\chtoh \ spectrum at selected positions.  The channel maps corresponding to the
lines 2$_0$--1$_0$ A$^+$ and 2$_{-1}$--1$_{-1}$ E are shown in
Fig.~\ref{fchch3oh}.

\begin{figure*}
\begin{center}
\includegraphics[scale=0.7,angle=0]{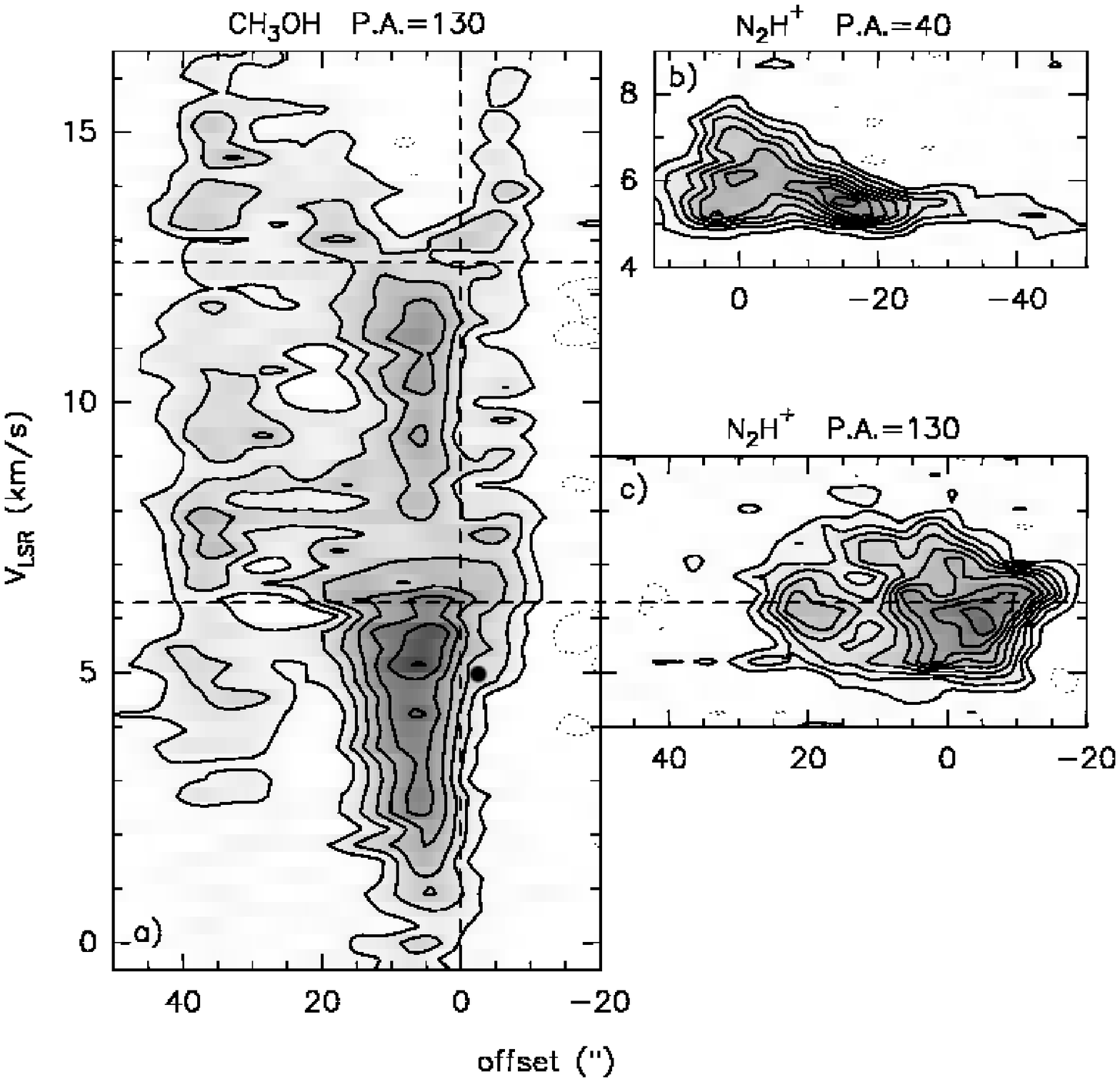}
\caption{
{\bf a)} \chtoh\ position-velocity (p-v) plot along outflow B
(P.A.$=$130\degr). Channel maps have been convolved with a beam of $5''\times
2''$, with P.A. perpendicular to the direction of the cut. Contours start at
0.17~Jy~beam$^{-1}$, and increase in steps of 0.17~Jy~beam$^{-1}$.   The
bottom dashed line indicates the systemic velocity for line 2$_0$--1$_0$ A$^+$
(taken as the reference line), at 6.3~\kms. The top dashed line indicates the
systemic velocity for line 2$_{-1}$--1$_{-1}$ E, at 12.6 \kms. 
{\bf b)} \nth\ p-v plot for the $F_1=0$--1 hyperfine across
outflow B and the main cloud (P.A.$=$40\degr). Positive positions are
toward the northeast. Contours start at 0.2~Jy~beam$^{-1}$, increasing in steps
of 0.2~Jy~beam$^{-1}$. 
{\bf c)} \nth\ p-v plot for the $F_1=0$--1 hyperfine along outflow B
(P.A.$=$130\degr). The velocity scale is placed matching the velocity scale of
the \chtoh\ emission. Contours start at 0.15~Jy~beam$^{-1}$, increasing in
steps of 0.15~Jy~beam$^{-1}$. For \nth, we did not convolve the channel maps.
In all panels, the central position corresponds to the Northern Warm Spot (see
Table~\ref{tsources}), and is marked by a vertical dashed line in panel a. 
\label{fpvridgeB}
}
\end{center}
\end{figure*}

The strongest emission of \chtoh \ is found to the southeast of BIMA~1,
elongated in the southeast-northwest direction, and covering a spatial
extension of $\sim$55$''$ (0.6~pc). The emission has a fork-like structure
(this is well observed  in the 7.7~\kms\ velocity channel of
Fig.~\ref{fchch3oh}), and extends through several channels.  This structure is
associated with high-velocity SiO\,(2--1) and CO\,(2--1) emission of outflow B
(Beuther \et\ 2004a). The position-velocity (p-v) plot  along  outflow
B is shown in Fig.~\ref{fpvridgeB}a. The most prominent feature in the plot is
a blue wing, $\sim$6~\kms\ wide, spanning from $0''$ to $15''$ offset
positions.  At negative offsets the \chtoh\ emission is dominated by a weak red
wing (this is better seen for the 2$_{-1}$--1$_{-1}$ E line). Farther away than
$\sim$25$''$ (which corresponds to the southeasternmost clump of
 outflow B), the emission has contributions from both redshifted and
blueshifted emission. This change of behaviour spatially coincides with BIMA~4,
located in the p-v plot at an offset position of 20$''$.

A clump to the east of  outflow B that can be seen in the integrated
intensity map shows narrow lines and appears only at systemic velocities (see
Figs.~\ref{fspecs} and \ref{fchch3oh}). We label this clump, which has not been
previously detected, 'methanol eastern clump'.

\begin{figure*}[htb]
\begin{center}
\includegraphics[scale=0.6,angle=0]{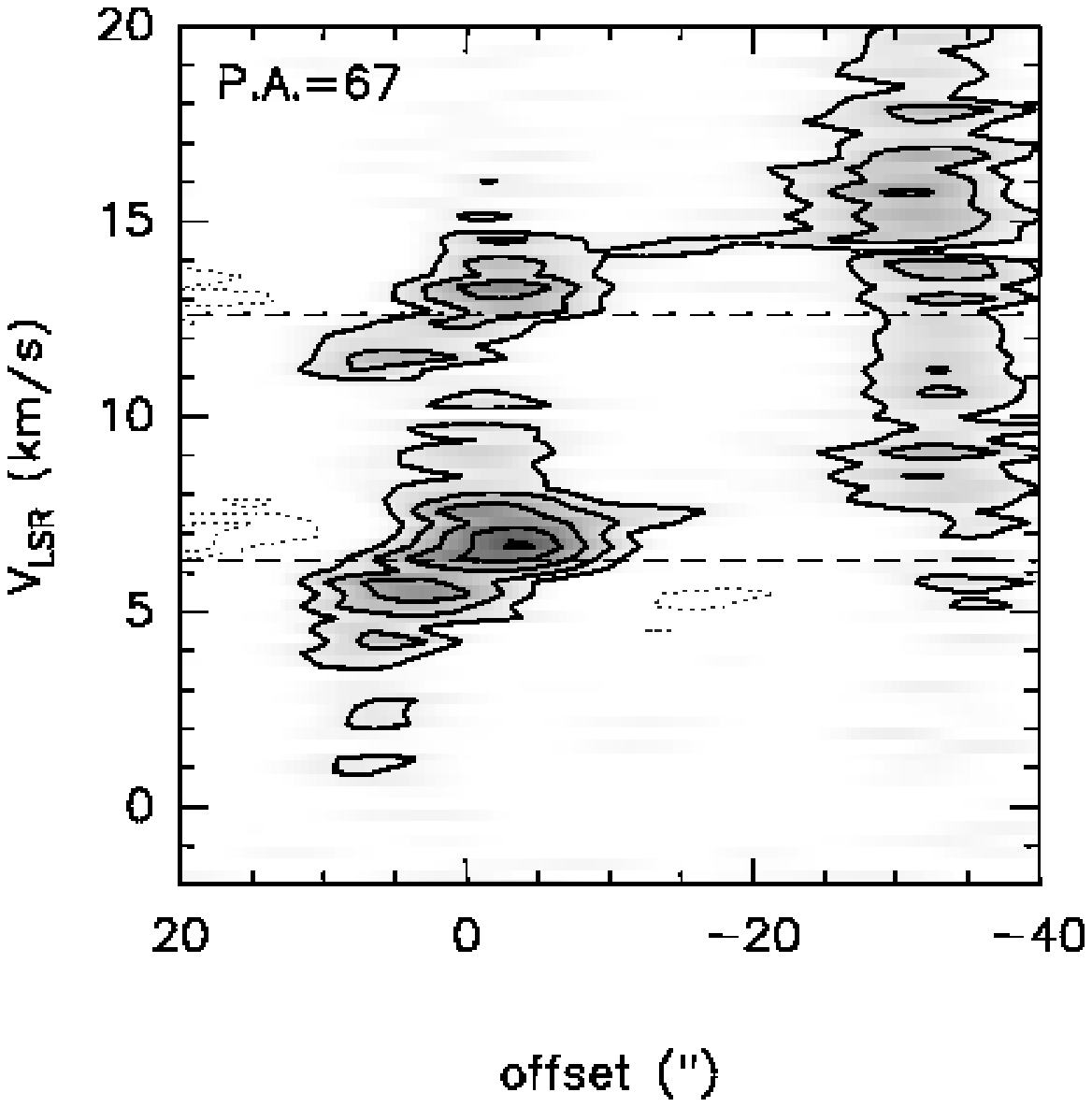}
\caption{
\chtoh\ p-v plot along outflow A, with P.A.$=$67\degr. Contours are $-0.25$,
$-0.15$, 0.15, 0.25, 0.35, 0.45, and  0.52~Jy~beam$^{-1}$. The velocity
resolution is 0.30 \kms, and the channel maps have been convolved with a beam
of $10''\times 5''$, with a P.A. perpendicular to the direction of the cut, in
order to recover the maximum  emission in each position.  The velocity range
includes two lines: 2$_0$--1$_0$ A$^+$ (at 6.3~\kms, bottom dashed line) and
2$_{-1}$--1$_{-1}$ E (at 12.6~\kms, top dashed line). The zero position
corresponds to mm1, and positive values go to the northeast.
\label{fpvridgeA}
}
\end{center}
\end{figure*}

In addition, there are two redshifted clumps, one associated with BIMA~1, and
the other close to the \uchii\ region (see, \eg\ velocity channels from 13 to
16~\kms), which probably are part of the redshifted CO\,(2--1) lobe of outflow
A (Beuther \et\ 2004a). We label the clump close to the \uchii\ region
'methanol western clump'. In Fig.~\ref{fpvridgeA}, showing the p-v plot along
 outflow A, we find that there is a velocity gradient, from 4~\kms\ at
$7''$, to 8~\kms\ at $-10''$, clearly seen in the two lines. This velocity
gradient is centered  around the zero offset position (corresponding to mm1).
The methanol western clump, seen at $\sim -35''$, is highly redshifted (line
2$_0$--1$_0$ A$^+$ is at $\sim$9~\kms) and shows broad red wings, spreading
$\sim$5~\kms.

\subsection{N$_2$H$^+$}

The zero-order moment map integrated for all the hyperfine transitions is
presented in Fig.~\ref{fm0}b. Figure~\ref{fspecs} shows the \nth\,(1--0) spectra
at selected positions.

The integrated \nth \ emission consists of a main cloud and a smaller and
weaker cloud to the west of the main cloud (western cloud). The size of the
main cloud and the western cloud are $\sim$50$''$ (0.5~pc), and  $\sim$15$''$
(0.15~pc), respectively. The four BIMA sources are located in the main cloud
with BIMA~1 close to the \nth\ emission peak. \nth \ emission is marginally
detected towards the \uchii\ region.

\begin{figure*}
\begin{center}
\includegraphics[scale=0.7]{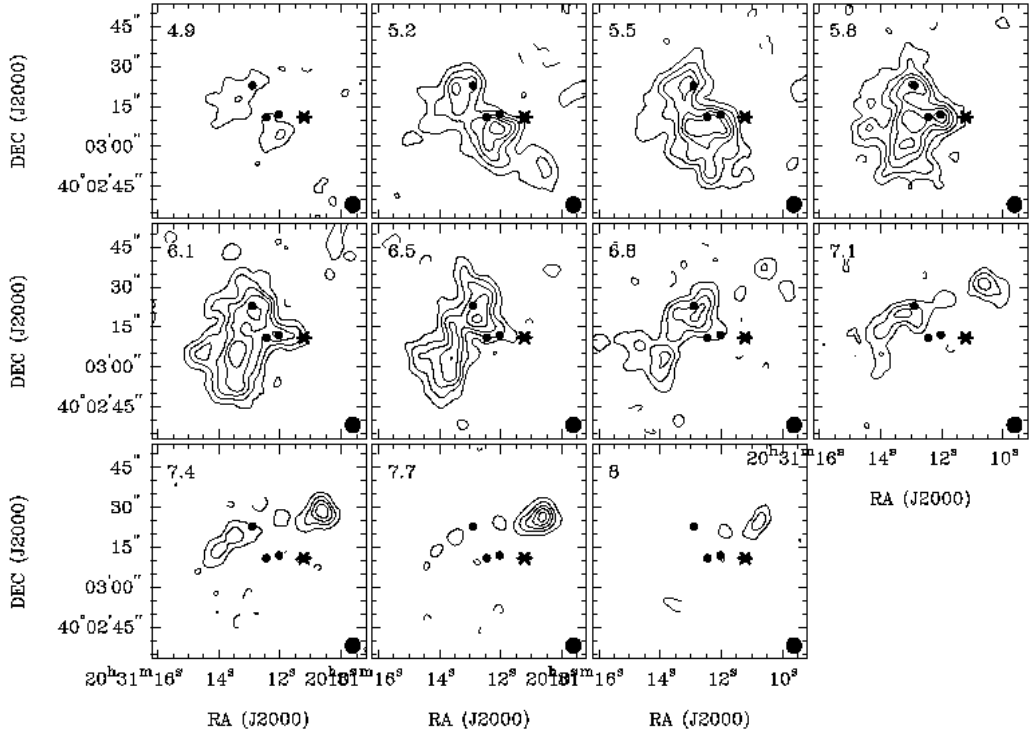}
\caption{
\nth\,(1--0) channel maps for the $F_1=0$--1 hyperfine toward IRAS~20293+3952.
Contour levels are $-4$, 4, 8, 12, 16, and 20 times the rms noise of the maps
0.07~Jy~beam$^{-1}$. Filled circles are the compact millimeter sources detected
by Beuther \et\ (2004b), and the star marks the position of the \uchii\ region.
The synthesized beam is shown in the bottom right corner. \label{fchn2h}
}
\end{center}
\end{figure*}

\begin{figure*}
\begin{center}
\includegraphics[scale=0.6]{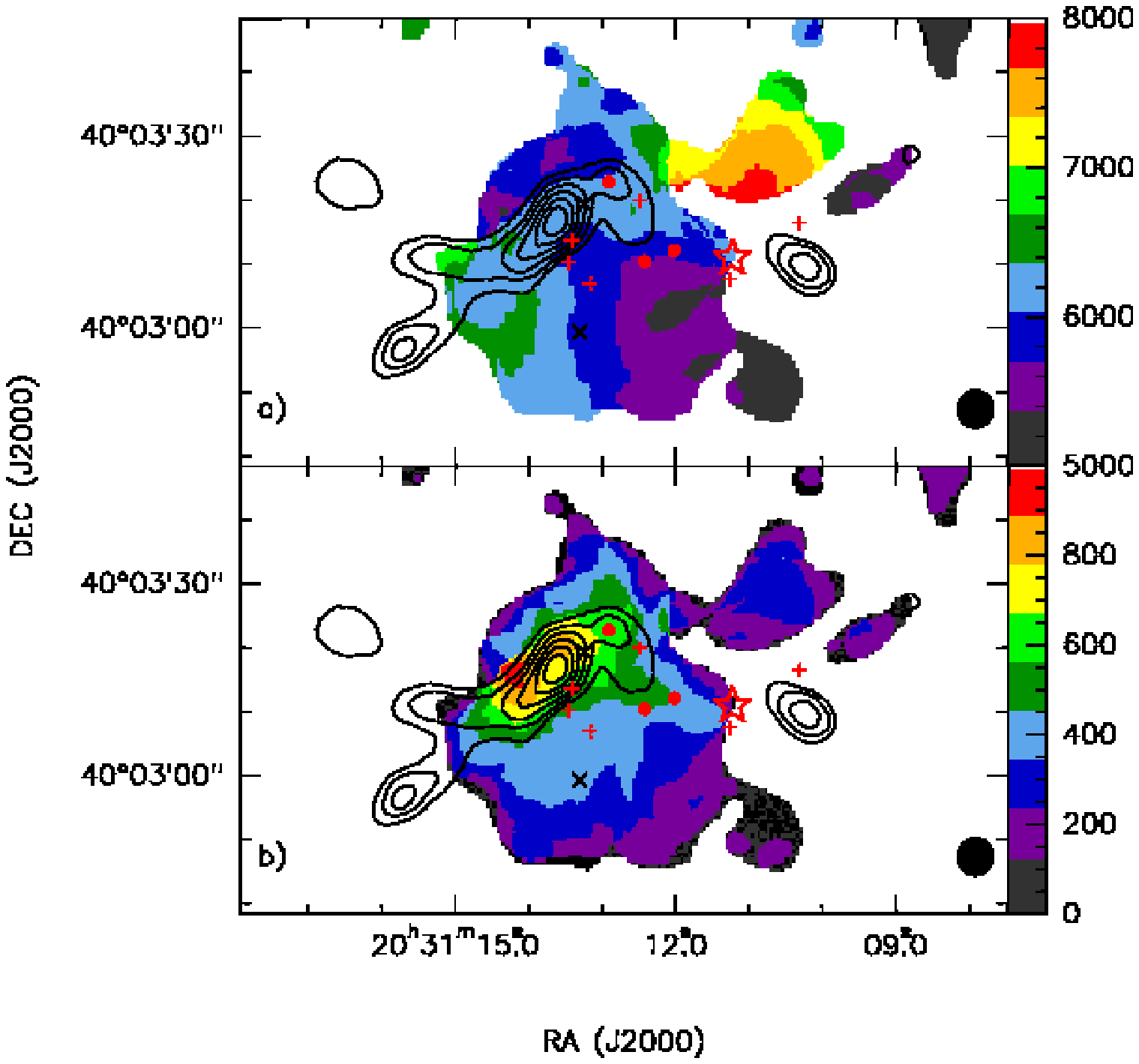}
\caption{
{\bf a)} Color scale: first-order moment map for the hyperfine $F_1=0$--1
line of \nth\,(1--0) toward IRAS~20293+3952.  
{\bf b)} Color scale: second-order moment map for the hyperfine $F_1=0$--1
line of \nth\,(1--0). In both figures black contours are the same as Fig.~\ref{fm0}a,
showing the \chtoh\,(2--1) emission, with contours starting at 13\%, and
increasing in steps of 15\% of the peak intensity. Color scales are in
m~s$^{-1}$. The synthesized beam is shown in the bottom right corner, and symbols
are the same as in Fig.~\ref{fm0}. Note that the second-order moment gives the
velocity dispersion, and must be multiplied by the factor
$2\sqrt{2\mathrm{ln}2} \simeq 2.35$ to obtain FWHM line widths.\label{fn2hm1m2}
}
\end{center}
\end{figure*}

The hyperfine $F_1=0$--1 was used for the analysis of the kinematics of the
\nth \ emission because it is not blended with the other hyperfines. In
Figs.~\ref{fchn2h} and \ref{fn2hm1m2} we show the channel maps and the first
and second-order moments for this hyperfine. The channel with maximum
intensity was found at the systemic velocity (6.3 \kms).  

The first-order moment map shows a small velocity gradient in the main cloud
with increasing velocities roughly from the southwest to the northeast. In
addition,  there is a filament spatially coincident with outflow B
(clearly visible in the 7.1 to 7.7~\kms\ channel maps of Fig.~\ref{fchn2h}),
which has very broad lines associated, as seen in the second-order moment map
(Fig.~\ref{fn2hm1m2}b). The p-v plot made across  outflow B and within
the main cloud (Fig.~\ref{fpvridgeB}b) shows the aforementioned velocity
gradient in the main cloud and the line broadening toward  outflow B
(at offset position zero). The p-v plot along  outflow B
(Fig.~\ref{fpvridgeB}c) shows clear line broadening all along the
outflow, but without the wing emission component found in the \chtoh\
emission.

The western cloud appears redshifted, at $\ge 7$ \kms, and has narrow lines
(Fig.~\ref{fchn2h}).  At these velocities, there is weak emission connecting
the western cloud with  outflow B. Figure~\ref{fpvwestcloud} shows the
p-v plot along a cut at P.A.$=$110\degr, picking up partially the outflow and
the western cloud.  From this plot, it seems that the western cloud extends 
$\sim$1$'$ (0.6~pc) toward the east, intersecting the main cloud.


\subsection{NH$_3$}

The zero-order moment map of the \nh\,(1,1) emission (integrated including the
main line and the inner satellites) is shown in Fig.~\ref{fm0}c. The overall
structure of \nh\,(1,1) resembles roughly the emission of the dust and the
emission of \nth. \nh\,(1,1) emission is also associated with the main cloud and
the western cloud. There is another \nh\,(1,1) clump to the northwest (not shown
in Fig.~\ref{fm0}), which is very weak in \nth \  (in part due to the primary
beam attenuation). A difference between \nth \ and \nh\,(1,1) emission is that
whereas \nth\ peaks close to the millimeter continuum sources, the strongest  \nh\,(1,1)
emission in the main cloud lies in the southeastern part. In addition, the
strongest \nh\,(1,1) emission over the entire region comes from the western
cloud, not from the main cloud. \nh\,(1,1) emission toward the \uchii\ region is
weak.

Figure~\ref{fm0}d  shows the zero-order moment map of the \nh\,(2,2) emission.
\nh\,(2,2) emission resembles closely that of the \nh\,(1,1). However, the
main difference with \nh\,(1,1) is that the strongest emission of the
\nh\,(2,2) in the entire region is found very close to BIMA~1. The \nh\,(2,2)
emission in the western edge of the main cloud shows an extension that
encompasses the position of the \uchii\ region.

Note that the dense gas traced by \nh\ and \nth\ is morphologically very
different from the gas traced by \chtoh.

\section{Analysis}

\subsection{Dust \label{sad}}

We compared the position of BIMA~1 with the millimeter source found by Beuther \et\ 
(2004b) at 3.5 mm at similar angular resolution. The peak of BIMA~1 is shifted
$\sim$1$''$ to the west of mm1.  The values of the peak intensities  (Table
\ref{tcont}) are in very good agreement with Beuther \et\ (2004a), and the flux
densities from Table \ref{tcont} are about 50\% larger than those reported in
Beuther \et\ (2004a).  The offset in positions and the larger flux densities
detected in this work could be produced by the different spatial filtering of
the BIMA and PdBI arrays, with the BIMA array more sensitive to large-scale
structures.


The continuum emission at 3.15 mm is probably due to thermal dust emission for
most positions of the region, since free-free emission from ionized gas, 
traced by the continuum emission at 3.6 cm (Fig.~\ref{fcontr}), is detected
only at the western edge of the main cloud, at the position of the \uchii\
region. However, Beuther \et\ (2004b) find a spectral index between 1.3 and
2.6~mm toward mm1a (the subcomponent to the west of mm1) of 0.8, and suggest
that the collimated ionized wind from outflow A could account for such a
spectral index. The contribution of the possible free-free emission from mm1a
to BIMA~1 is small, given that the intensity of mm1 is five times larger than
the intensity of mm1a, and thus essentially all continuum emission at 3~mm is
due to thermal dust emission.


To derive masses from the flux densities at 3.15~mm, we corrected the
rotational temperature, between 15 and 25~K (calculated from \nh\,(1,1) and
\nh\,(2,2), see \S~\ref{satrot}), to kinetic temperature (Walmsley \&
Ungerechts 1983), by following Tafalla \et\ (2004), yielding kinetic
temperatures in the range 17--34~K. The masses (Table~\ref{tcont}) are 2 times
larger than those obtained by Beuther \et\ (2004b) from observations at 3.5~mm.
The difference arises from the fact that BIMA is detecting more flux than PdBI
observations, and from the different dust temperatures assumed. While Beuther
\et\ (2004a) assume a dust temperature (derived from a graybody fit to the
spectral energy distribution in the entire region) of 56~K, we adopted the
kinetic temperatures obtained from the \nh\ observations,  which allowed us to
separate the contribution from each source, thanks to the high angular
resolution. The lower temperatures adopted imply larger masses in order to
produce the same flux density.


\subsection{Rotational temperature and column density maps \label{satrot}}

We obtained \nth\ and \nh\ spectra for positions in a grid of $2''\times 2''$
in the main cloud and the western cloud, and  fitted the hyperfine structure of
each spectrum, or a single Gaussian for \nh\,(2,2).  We performed the fits only
for those spectra with an intensity greater than 5$\sigma$ for \nh\,(1,1),
and greater than 4$\sigma$ for \nh\,(2,2). We set a higher cutoff for
\nh\,(1,1) to ensure that not only the main line is detected but also the
satellites. The results from the fits indicate that the entire main cloud is
essentially optically thin for \nth\  ($\tau_\mathrm{TOT} \le 1.5$), but
optically thick for \nh\,(1,1) ($\tau_\mathrm{TOT} \le 12$). In both cases, the
highest opacities are reached in the southeastern side of the cloud, around
BIMA~3.

\begin{table*}{\small
\caption{Significant local temperature enhancements \label{ttrot}}
\centering
\begin{tabular}{ccccc}
\hline\hline
\multicolumn{2}{c}{Position}
&$\Trot$
&$\Delta T$$^\mathrm{a}$
&Possible
\\
$\alpha (\rm J2000)$
&$\delta (\rm J2000)$
&(K)
&(K)
&Counterparts
\\
\hline
20:31:13.26
&40:02:58.9
&$20\pm2$
&$7\pm2$
& 2.12 $\mu$m faint emission (SWS$^\mathrm{b}$)
\\
20:31:13.23 
&40:03:19.5
&$29\pm6$
&$12\pm6$
& 2.12 $\mu$m faint emission (NWS$^\mathrm{b}$)
\\
20:31:13.44
&40:03:13.4
&$24\pm3$
&$7\pm3$
& IRS~5
\\
20:31:14.96
&40:03:03.4
&$23\pm4$
&$9\pm4$
&H$_2$ knot c~$^\mathrm{c}$
\\
\hline
\end{tabular}
\begin{list}{}{}
\item[$^\mathrm{a}$] We assumed that each local maximum can be described by a
'gaussian + background level' model, with the gaussian width and position angle
fixed and equal to the synthesized beam. The temperature enhancement, $\Delta
T$, is defined as the difference between the local maximum and the background
level. We considered as significant the temperature enhancements with $\Delta T
\geq 2\sigma$, with $\sigma$ being the uncertainty.
\item[$^\mathrm{b}$] NWS: Northern Warm Spot; SWS: Southern Warm Spot
\item[$^\mathrm{c}$] from Kumar \et\ (2002).
\end{list}
}
\end{table*}

From the results of the fits to the \nh\,(1,1) and the \nh\,(2,2) spectra  we
computed the rotational temperature and \nh \ column density maps.   We derived
the rotational temperature following the standard procedures (Ho \& Townes
1983; Sep\'ulveda 1993), and assuming  that the excitation temperature and the
intrinsic line width were the same for both \nh\,(1,1) and \nh\,(2,2). For the
\nh \ column density derivation, we followed the expression given in Table 1 of
Anglada \et\ (1995), and assumed that the filling factor was 1.



The map of the rotational temperature is presented in Fig.~\ref{fpxpx}a.  The
maximum value, $38 \pm 15$~K, is  reached at the position of the \uchii\ region. 
Interestingly, around the apparent dense gas cavity  west to BIMA~1, \nh\ shows
high temperatures, with a maximum of $34 \pm 9$~K (the fits to the \nh\,(1,1)
and \nh\,(2,2) at these positions are reasonably good, so this heating must be
real). Toward the western cloud we find that temperature is essentially
constant, around 16~K, slightly decreasing toward the center of the cloud.

There are some local maxima of rotational temperature in the map. In
order to test the significance of these local maxima, we fitted them with a
'gaussian + background level' model in a region a few times the synthesized
beam size, with the gaussian width and position angle fixed and equal to the
synthesized beam. We defined the temperature enhancement as the difference
between the local maxima and the background level. The rms of the residual
image in the fitted region is $<1.5$~K, and the positions of the gaussian peaks
are within $0\farcs2$ of the local maxima positions. In Table~\ref{ttrot} we
list the positions of the significant local maxima, their temperature, and
temperature enhancement. We considered that a local maxima is significant when
the temperature enhancement is at least two times the uncertainty. There is a
local maximum that is nearly coincident (within $0\farcs5$) with IRS~5, with a
value of $24 \pm 3$~K. The warmest local maximum is located  $\sim$5$''$ to the
southeast of mm1 (labeled as 'Northern Warm Spot'), reaching $29 \pm 6$~K, and
is apparently associated with faint continuum emission at 2.12 $\mu$m (see
Fig.~\ref{fcontr}). There is another local maximum associated also with faint
emission at 2.12 $\mu$m, which is found  $10''$ to the southwest of BIMA~3,
with $20 \pm 2$ K (labeled as 'Southern Warm Spot'). Another local maximum is
associated with the H$_2$ knot c (Kumar \et\ 2002), lying on the axis of
outflow B.  Finally, note that there is also some heating about $20''$ to the
southeast of the Northern Warm Spot, but we did not consider this heating in
the table because it falls in the edge of the region where we fitted the
spectra.  Note that all the temperature enhancements in
Table~\ref{ttrot} have infrared emission associated, giving support to their
significance.



Figure~\ref{fpxpx}b shows the resulting column density map for \nh, corrected
for the primary beam response. An obvious feature from the \nh\ column density
map is that the highest values are found to the southeast of the main cloud,
where we found the lowest values in the rotational temperature map
(Fig.~\ref{fpxpx}a). In the northern part of the cloud the column density is
higher around the Northern Warm Spot.  It is worth noting that the \nh \ column
density decreases toward IRS~5.  Note also that in the western cloud the column
density increases toward the center.

We calculated the \nth \ column density by following Benson \et\ (1998), 
taking into account the opacity effects, and correcting for the primary beam
attenuation. In the expression for the column density, the value for $\Tex$ was
derived from the hyperfine fit made with CLASS, and assuming a filling factor
of 1. The resulting map is shown in Fig.~\ref{fpxpx}c.  Contrary to \nh, the
map of the \nth \ column density has the maximum value very close to the
position of the Northern Warm Spot and BIMA~1, and not in the southern side of
the main cloud.


\subsection{The \nh/\nth\ column density ratio map \label{saratio}}

In order to compare the \nh \ emission with the \nth \ emission, we convolved
the  \nh \ and \nth \ channel maps to obtain a final beam of $7''$ (the major
axis of the \nh \ and \nth \ beams). We fitted the spectra in each position of
a grid of $4'' \times 4''$ in the convolved maps, and derived the column
density for \nh \ and \nth \ following the same procedures outlined above, and
correcting for the primary beam of each interferometer. We computed the
\nh/\nth \ column density ratio map, and the result is shown in
Fig.~\ref{fpxpx}d. From the \nh/\nth \ ratio map,  one can see a clear gradient
from the northwest  of the main cloud, with a ratio around 50, to the
southeast, where the ratio reaches values up to $\sim$300. Such high values are
also reached in the western cloud.

\section{Discussion}

\subsection{General properties of the dense gas \label{sddensegas}}

\subsubsection{Rotational temperature}

The temperature distribution in the main cloud can be clearly separated into
two parts: the northern side, with an average temperature of $\sim$22~K,  and
the southern side, with an average temperature of $\sim$16~K.  It is
interesting to note that almost all the YSOs in the region (\S~\ref{sdsources})
are associated with the northern side of the cloud, while in the cold southern
side we found very few hints of active star formation. Thus, the higher
temperature in the northern side is probably due to internal heating from the
embedded YSOs. As for the southern side, the average temperature of 16~K is
higher than typical temperatures for low-mass externally heated starless cores
($\sim$10~K, \eg\ Tafalla \et\ 2002, 2004). Such higher temperatures could
indicate that there are low-mass non-detected YSOs heating the southern side of
the main cloud, or that there is external heating. A similar result was found
by Li \et\ (2003), who derived temperatures of $\sim$15~K toward massive (with
masses similar to the mass derived for the main cloud, see next paragraph)
quiescent cores, with no signs of star formation, in Orion. Li \et\ (2003) find
that the temperatures of the massive quiescent cores can be well explained by
the dust being heated by the external UV field from the Trapezium, at a
distance of $\sim$1~pc from the massive cores. In our case, the B1 star
associated with the \uchii\ region (see \S~\ref{sdsources}) could be the source
of heating of the southern side of the main cloud, at $\sim$0.5~pc from the
\uchii\ region. However, we find no temperaure gradient across the south of the
main cloud. Alternatively, the heating could be produced by  the nearby O and B
stars of the Cygnus~OB2 association (Le Duigou \& Kn\"odlseder 2002). 



\subsubsection{Column density and mass}

The total mass of gas and dust derived from the dust continuum emission at
3.15~mm, integrated in a region including all the BIMA sources, is
$\sim$100~\mo \  (assuming $\Td=25$ K, and $\beta=1.5$), about half of the
total mass from dust at 1.2~mm observed with the IRAM~30\,m by Beuther \et\
(2002a, 2005a), 230~\mo. We compared this mass with the mass derived from the
high-density tracers. The average \nh\ column density for the main cloud,
1.45$\times 10^{15}$~cm$^{-2}$, is similar to the values obtained by Jijina
\et\ (1999) for starless regions, and the derived average \nth\ column density
of 1.21$\times 10^{13}$~cm$^{-2}$ agrees with the \nth\ column densities
obtained by Pirogov \et\ (2003) for molecular cloud cores with massive stars
and star clusters.  The mass estimated for the main cloud from \nh\ is
$\sim$340~\mo, assuming an abundance for \nh\ of 10$^{-8}$, and the mass
derived from the \nth\ column density map is 75--190~\mo\ for \nth\ abundances
in the range (2--5)$\times10^{-10}$ (Pirogov \et\ 2003; Tafalla \et\ 2004). 
Thus, although the uncertainty in the abundances is certainly affecting the
estimation of the mass from \nh\ and \nth\  (indeed, there is strong chemical
differentiation across the main cloud, see \S~\ref{saratio}), we obtained
similar masses from the dense gas tracers and the dust emission. Regarding the
western cloud, the derived mass from \nh\ and \nth\ lies in the range
11--47~\mo.




\subsubsection{Starless core candidates \label{sddensegasstarless}}

We identify three starless core candidates in the region: BIMA~3, BIMA~4, and
the western cloud. For these three cores, the \nh\ column density is high,
$\gtrsim 3 \times 10^{15}$~cm$^{-2}$, the rotational temperature is low,
$\sim$14~K, there is no near-infrared emission in the $J$, $H$, and $K$ bands
of 2MASS, no mid-infrared emission in the 24~$\mu$m MIPS (Multiband Imaging
Photometer for \emph{Spitzer}\footnote{see
http://ssc.spitzer.caltech.edu/mips/}) band, and no signs of molecular outflows
emanating from the cores (see \S~\ref{sdinter} for a discussion on the outflow
emission in BIMA~4). Obviously, these cores could harbor low-mass YSOs
(with masses $\lesssim\!1$~\mo, as estimated from the available data)
non-detected at millimeter and infrared wavelengths. However, the rotational
temperature map of Fig.~\ref{fpxpx}a shows local minima associated with these
cores, specially with BIMA~4 and the western cloud, supporting that the heating
is external and thus that there is no embedded YSO yet.

It is worth noting that, for the three cores, line widths derived from \nh\
(and also \nth) hyperfine fits are around 1~\kms\ (Table~\ref{tsources}),
which are considerably larger than typical line widths measured toward
starless cores in low-mass star-forming regions, $\sim$0.2~\kms\ (see, \eg\
Crapsi \et\ 2005). For the massive quiescent cores studied by Li \et\ (2003)
the \nh\ line widths were between 0.6 and 0.9~\kms, similar to what we find.
Since the thermal component is around $\sim$0.2~\kms\ at 15~K, the line widths
that we found suggest that these cores are dominated by turbulence, as Li \et\
(2003) state, or that there are important unresolved ($\la 0.05$~pc)
systematic motions. 




\subsubsection{Effects of chemical abundance and density}

The \nh/\nth \ ratio map of Fig.\ \ref{fpxpx}d shows a clear gradient in the
ratio across the cloud, from northwest, with low values, to southeast, with
the highest values. In particular, there seems to be an anticorrelation
between the \nh/\nth \ ratio and the evolutionary stage of the sources
embedded in the cloud. On one hand, in the north of the main cloud we found
most of the YSOs in the field, where the value of the ratio is $<100$. On the
other hand, the ratio rises up to $\sim$300 in the southeastern side of the
main cloud and in the western cloud, where we found starless core candidates.
The values of the ratio in the entire region are consistent with those derived
for low-mass star-forming regions by other authors (Caselli \et\ 2002; Hotzel
\et \ 2004), who also find the same trend of low values (about 60--90)
associated with YSOs, and high values (140--190) associated with starless
cores. 

One could consider whether the high values for the \nh/\nth \ ratio in the
southeastern side of the main cloud and in the western cloud are just
reflecting the different excitation conditions of  \nh\ and \nth, since the
critical density of \nh, $\sim$10$^4$~\cmt\ (\eg\ Ho 1977), is  lower than the
critical density of \nth, $\sim$10$^5$~\cmt\ (\eg\ Joergensen 2004), and thus
in regions where the density is $\sim$10$^4$~\cmt, \nh\ could be  thermalized
while \nth\ is faintly detected because not thermalized. We did a rough
estimation of the density in the southeastern side of the main cloud from the
continuum emission observed at 3.15~mm toward BIMA~3, adopting the mass derived
in this work (see Table~\ref{tcont}) and a size of $\sim$10$''$, and obtained a
volume density of $\sim$10$^5$~\cmt.  For this density, the transitions
of both \nh\ and \nth\ are thermalized enough to be detectable. Furthermore, as
the effects of opacity and different $\Tex$ in the cloud were already taken
into account in the calculation of the column densities, the different
excitation conditions of \nh\ and \nth\ cannot play an important role in the
large gradient of one order of magnitude in the \nh/\nth\ ratio across the main
cloud.


\begin{figure}
\begin{center}
\includegraphics[scale=0.7,angle=0]{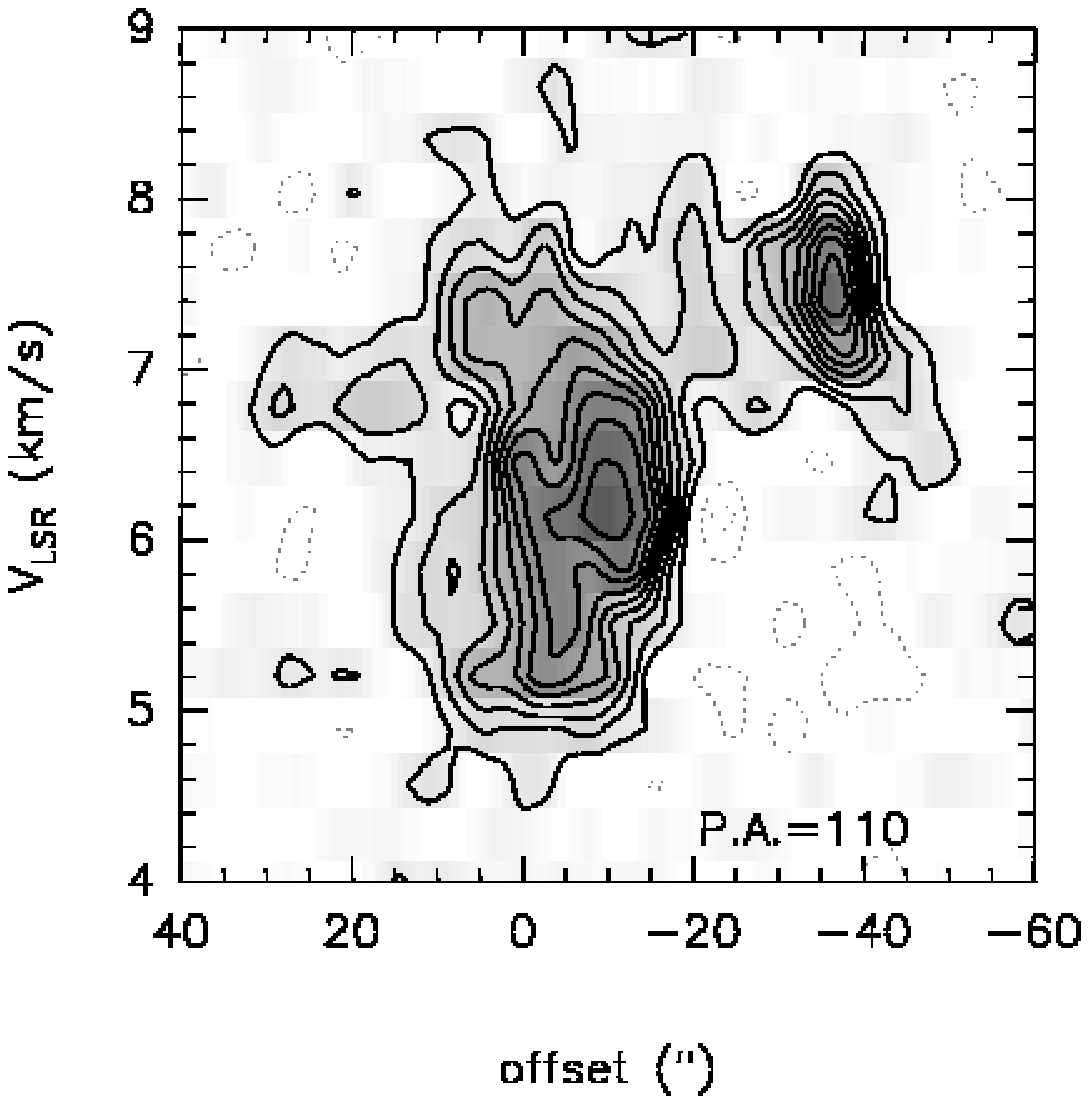}
\caption{
\nth\ p-v plot for the $F_1=0$--1 hyperfine at P.A.$=$110\degr, that is, along
the western cloud and the northern side of the main cloud. Central position is 
$\alpha(J2000)=20^{\mathrm h}31^{\mathrm m} 13\fs63$;
$\delta(J2000)=+40\degr03\arcmin16\farcs7$ (methanol peak), and positive 
position offsets are toward southeast. Contours start at 0.15~Jy~beam$^{-1}$,
increasing in steps of 0.15~Jy~beam$^{-1}$. \label{fpvwestcloud}
}
\end{center}
\end{figure}

\begin{figure}
\begin{center}
\includegraphics[scale=0.85]{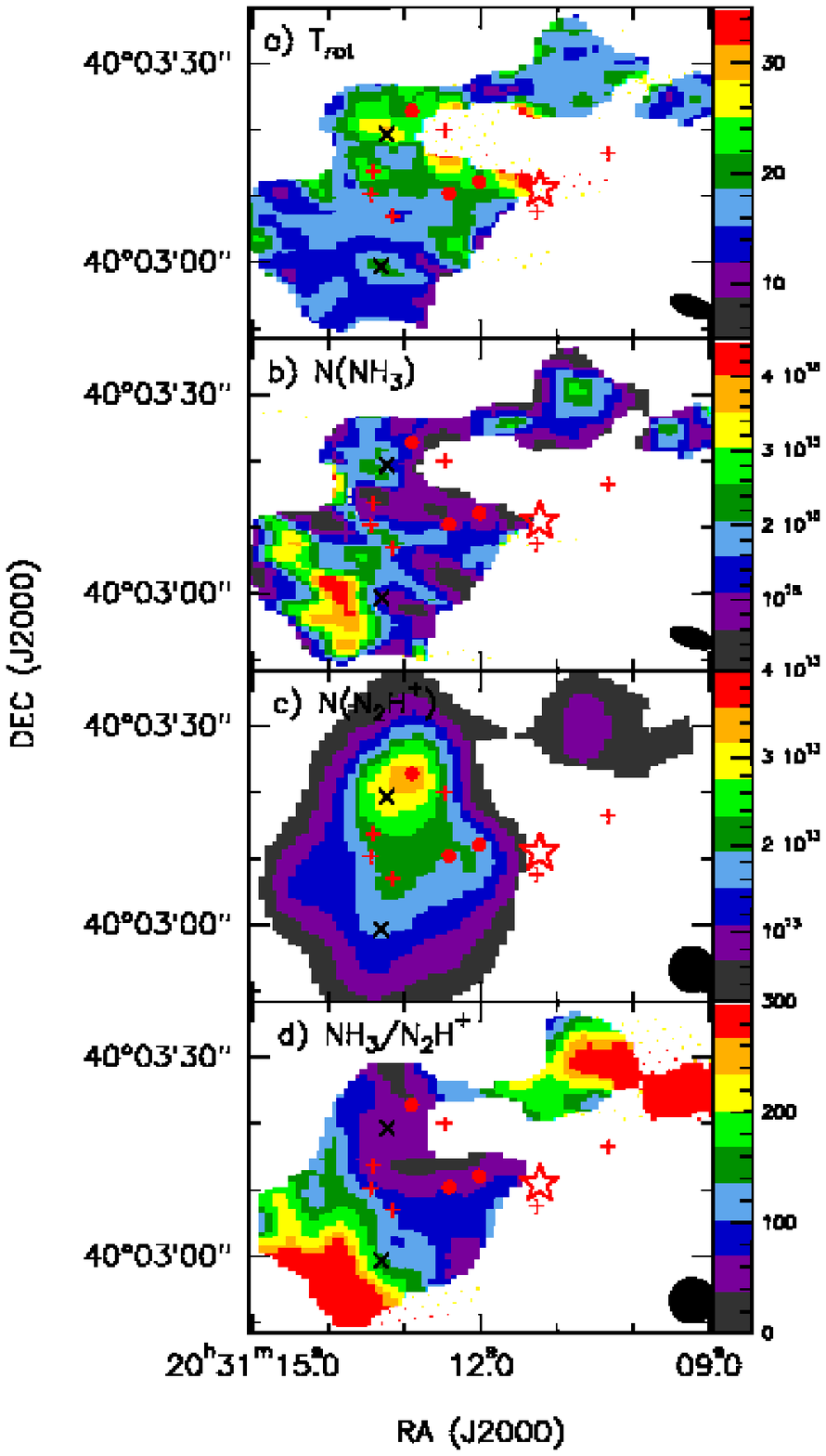}
\caption{
{\bf a)} Rotational temperature map from \nh\,(1,1) and \nh\,(2,2) obtained for
the main cloud and the western cloud of the IRAS~20293+3952 region. Scale units
are in K.
{\bf b)} \nh \ column density map. Scale units are in cm$^{-2}$.
{\bf c)} \nth \ column density map convolved to an angular resolution of
$7''$. Scale units are in cm$^{-2}$.
{\bf d)} N(\nh)/N(\nth) ratio map after convolving the \nh\ column density map
from panel b to a final angular resolution of $7''$. Symbols are the same as
in Fig.~\ref{fm0}. Beams are shown in the bottom right corner, and are
$6\farcs9 \times 3\farcs0$, with P.A.$=71\fdg5$ for panels a and b, and $7''$
for panels c and d.
\label{fpxpx}
}
\end{center}
\end{figure}

A possible explanation for the higher \nh/\nth \ ratio toward starless cores
may be given in terms of abundance and density. For long-lived
($\sim$10$^6$~yr) cores with moderate densities ($10^4$--$10^5$ \cmt), there is
a 'golden period' of high abundance for \nh\  (Aikawa \et\ 2005; S.
Viti, private communication), while \nth\ keeps a constant abundance, as has
been seen toward starless cores (\eg\ Tafalla \et\ 2004). This is consistent
with the \nh/\nth \ ratio being high in the southern side of the main cloud,
where we find starless cores with faint dust emission, indicating that the
volume density is moderate. For higher densities ($\sim$10$^6$~\cmt) one may
consider two mechanisms to explain the low \nh/\nth\ ratio.  On one hand, the
high-density cores, which have short free-fall timescales, may not have enough
time to produce an amount of \nh\ comparable to the southern (moderate-density)
cores. On the other hand, the \nh\ may start to freeze out onto dust grains,
while \nth\ seems to remain unaffected up to densities of $\sim$10$^7$ \cmt\
(Aikawa \et\ 2003,  2005). Note that the temperature in the main cloud
is on average low enough so that freezing out of \nh\ is possible (hot spots
are very localized). In the main cloud the highest densities are  reached in
the northern side, where there is the strongest dust emission. We did a rough
estimation of the average volume density toward BIMA~1, from the derived mass
and the size (at the 5$\sigma$ contour level) of BIMA~1 (Table~\ref{tcont}),
and obtained an average volume density toward BIMA~1 of $5 \times 10^5$~\cmt.
Consequently, we find low values of the ratio associated with  high volume
densities, and thus, with most of the YSOs in the main cloud.

\subsection{Interaction of the YSOs with the surrounding gas \label{sdinter}}

\subsubsection{The \uchii\ region:}

\begin{figure}
\begin{center}
\includegraphics[scale=0.7]{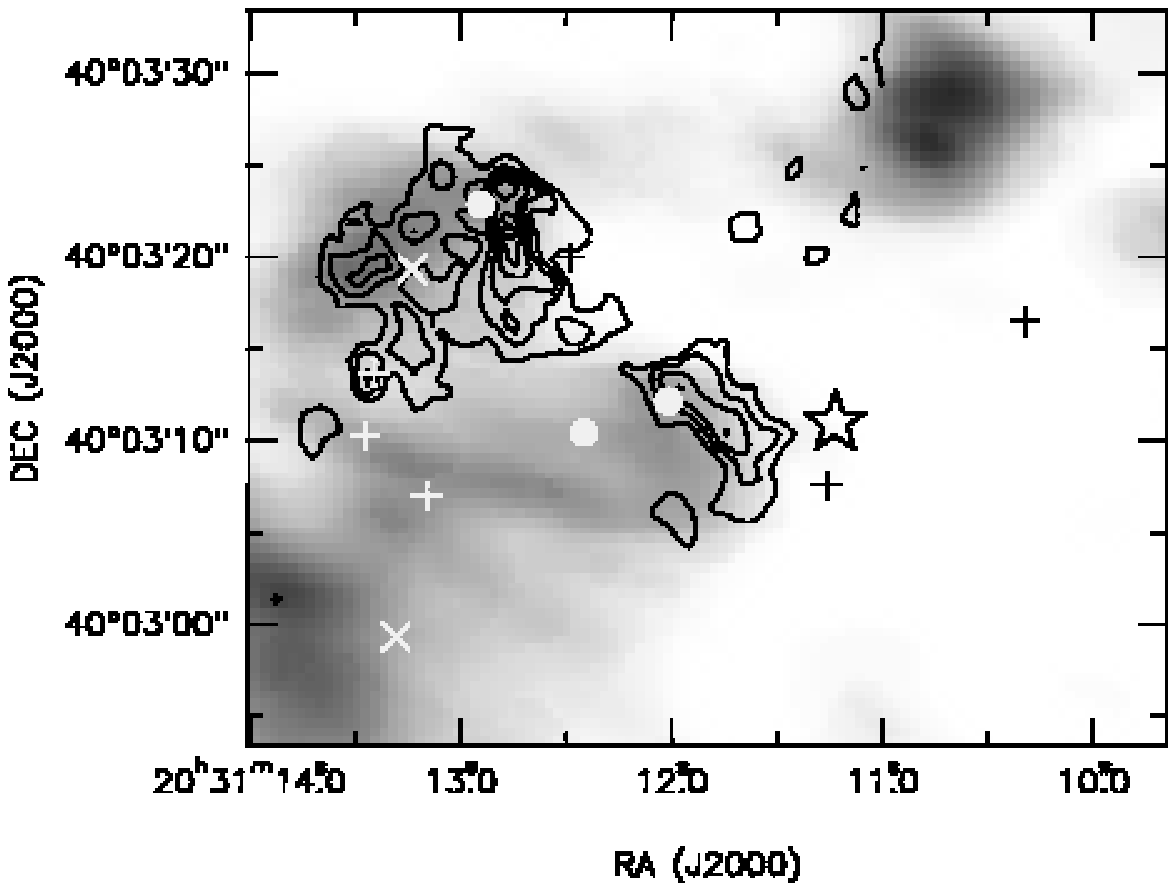}
\caption{
Zero-order moment map of the CN\,(1--0) emission (contours) from Beuther \et\
(2004b) in the IRAS~20293+3952 region, overlaid on the \nh\,(1,1) emission
shown in Fig.~\ref{fm0}c (grey scale). CN\,(1--0) emission has been integrated
from 4 to 12~\kms, and contours start at 15\% of the peak intensity,
273~Jy~beam$^{-1}$~\kms, increasing in steps of 15\%.  Symbols are the same as
in Fig.~\ref{fm0}. Beams are $6\farcs9 \times 3\farcs0$, with P.A.$=$72\degr\
for \nh\,(1,1), and $1\farcs46 \times 1\farcs21$, with P.A.$=$32\degr\ for
CN\,(1--0). 
\label{fcn}
}
\end{center}
\end{figure}

In Fig.~\ref{fcn} we overlay the CN\,(1--0) emission from Beuther \et\
(2004b) on the \nh\,(1,1) emission. The morphology of the CN emission next to
the \uchii\ region is following the edge of the dense cloud traced by \nh\,(1,1),
and is facing the \uchii\ region. CN emission is often found in photon-dominated
regions, tracing the part of the molecular cloud which is exposed to an intense
flux of FUV photons from a nearby young OB star (see, \eg \ Fuente \et \ 1993;
Simon \et \ 1997; Boger \& Sternberg 2005). In addition, we found heating
toward this edge of the cloud (Fig.~\ref{fpxpx}a), with rotational temperature
progressively increasing when approaching the \uchii\ region. All this indicates
that the \uchii\ region is radiatively interacting with the main cloud.  The
mechanical interaction of the compression front of the \uchii\ region with the
main cloud is likely traced by the H$_2$ emission at 2.12 $\mu$m (Kumar \et \
2002; Beuther \et\ 2004a), which reveals a ring-like structure around the \uchii\
region, of radius $\sim$8$''$, reaching the edge of the main cloud. 


\subsubsection{Outflow A:} 

Beuther \et\ (2004a) discovered a highly collimated outflow in CO\,(2--1)
(outflow A), elongated in the northeast-southwest direction, with mm1 being
its driving source. The axis of the outflow is shown in Fig.~\ref{fm0}.

The velocity gradient seen in \chtoh\ toward mm1 (see \S~\ref{srch3oh} and
Fig.~\ref{fpvridgeA}) is consistent with tracing material entrained by  ouflow
A. In addition,  the  p-v plot shows a clump at offset position $\sim -35''$,
the methanol western clump, with wide redshifted wings (see Figs.~\ref{fspecs}
and \ref{fpvridgeA}), and matching well the red lobe of CO\,(2--1) and the
redshifted clump of SiO\,(2--1) from Beuther \et\  (2004b). The methanol
western clump appears in the axis of outflow~A, just downstream of a weak dust
filament that connects IRS~1 and IRS~2 (see Figs.~\ref{fcontr} and \ref{fm0}a).
This suggests that the interaction of outflow A with the filament sweeps-up and
chemically alters the material of the filament, originating the methanol
western clump.


The red lobe of outflow A, which starts at mm1, propagates through the cavity
apparent in the \nth\ and \nh\ zero-order moment maps (see Fig.~\ref{fm0}). The
rotational temperature and column density maps of Fig.~\ref{fpxpx} show that
the walls of this cavity are characterized by remarkably high temperatures and
low column densities.  In addition, the \nh \ gas of the walls of the cavity is
redshifted (as the lobe of outflow A), and shows line broadening. Thus, this
cavity  could  have been  excavated by outflow A, and the high temperatures
could arise from the shock interaction of outflow A with the walls of the
cavity. This has been observed toward other outflows driven by high-mass stars
(\eg\ AFGL~5142: Zhang \et\ 2002).



\subsubsection{Outflow B:} 

Outflow B was first identified by Beuther \et\ (2004a) in SiO\,(2--1), which
shows a very strong blue lobe elongated in the northwest-southeast direction,
and in CO\,(2--1), with faint high-velocity blueshifted emission. \chtoh\
emission from outflow B is morphologically very similar to the SiO\,(2--1) blue
lobe and also splits up into two lobes  in a fork-like structure at the
position of BIMA~4. A scenario in which outflow~B is interacting with a
starless core, BIMA~4, naturally explains  the partial deflection of the
outflow seen in the SiO and \chtoh\ emission. There are several pieces of
evidence that support this scenario. First, the \chtoh\ emission splits up and
has a change in kinematics when it reaches BIMA~4 (\S~\ref{srch3oh}). Second,
at this position we found some heating and the \nh/\nth \  abundance ratio
shows lower values than those in the southern part of BIMA~4
(Fig.~\ref{fpxpx}a,d; temperature enhancement and chemical variations are
signatures of shocked molecular gas, \eg\ L1157: Bachiller \& P\'erez
Guti\'errez 1997). Third, the \nth\ line broadening along outflow B (which is
also observed in \nh) indicates that the outflow is already interacting with
the molecular cloud (Fig.~\ref{fn2hm1m2}b). Finally, downstream of BIMA~4,
there is heating (see Fig.~\ref{fpxpx}a) associated with the H$_2$ knot c of
Kumar \et\ (2002).   The partial deflection of an outflow due to the
interaction with a dense quiescent clump has been found toward other
star-forming regions (\eg\ IRAS~21391+5802, Beltr\'an \et\ 2002). Interaction
of an outflow with a dense flattened \nh\ core has been reported too toward
NGC\,2024 by Ho \et\ (1993), where the outflow sweeps the material off the
surface of the \nh\ core. Similarly, outflow B interacting with the BIMA~4 core
could be also sweeping the material off the surface of the core, producing all
the observational features described above. 


In addition, the properties of the methanol eastern clump  (located
$25''$ to the northeast of BIMA~4), which is found at ambient velocities, and
with narrow lines (around 1~\kms, Fig.~\ref{fspecs}),  could be a
consequence of the illumination by the UV radiation coming from the interaction
of outflow B with BIMA~4. The illumination of the UV radiation from shocks has
been proposed as the mechanism of enhancement of the emission of some species
in clumps ahead of shocks, in particular of \chtoh\ and \nh\  (see \eg\
Torrelles \et\ 1992b; Girart \et\ 1994, 2002). With the present observations,
we were not very sensitive to the \nh\ emission toward the methanol eastern
clump because the clump is located beyond the VLA primary beam FWHM (see
Fig.~\ref{fm0}c,d). Typically, such illuminated clumps in low-mass star-forming
regions have sizes of 0.05--0.1~pc, and are located at distances to the shock
of the order of 0.1 pc (Girart \et\ 1998). This is consistent with the methanol
eastern clump, which has a size of $\sim$0.15~pc, and is located about 0.2~pc
from BIMA~4.



It is interesting to note that the morphological features of the dense gas
around outflow B are different from outflow A. While outflow A has already
cleared up the dense molecular gas, creating a well defined cavity, outflow B
is probably in the embrionary phase of creating the cavity. This suggests that
possibly outflow B is younger or less energetic than outflow A, in agreement
with the measurements of Beuhter \et\ (2004a).  It is worth noting that
the line width of the dense gas associated with outflow B is significantly
higher than the thermal line width for both \nh\ and \nth\ ($\sim\!1.8$~\kms\
compared to $0.2$~\kms). A similar result is found by Wang \et\ (2006) toward
massive protostellar cores mapped in \nh\ with high angular resolution. Given
that we do not observe systematic motions in the dense gas associated with
outflow B, we suggest that the large line width in the dense gas is due to
turbulence injected by the passage of the outflow (see Fig.~\ref{fn2hm1m2}b).

Regarding the driving source of outflow B, a possible candidate is mm1, as
Beuther \et\  (2004b) proposed. If this was the case, mm1 would be a  binary
system of jet sources, since outflow A is clearly associated with mm1.  Another
possibility would be that BIMA~4 is the driving source of outflow B. However,
the kinematics of the gas as traced by \chtoh\ show that there is no clear
symmetry with respect to BIMA~4 (see Fig.~\ref{fpvridgeB}a: BIMA~4 is at an
offset position $\sim$20$''$). Another candidate of being the driving source of
outflow B is the Northern Warm Spot, which is aligned with outflow B, and is
located at the begining of the blue lobe of \chtoh. Furthermore, we found a
clear symmetry with respect to the Northern Warm Spot in the p-v plot of
Fig.~\ref{fpvridgeB}a (see \S~\ref{srch3oh}), strongly suggesting that the
Northern Warm Spot could be the driving source of outflow B. In
\S~\ref{sdsources} we estimate an associated mass of $\sim$0.7~\mo\ for the
Northern Warm Spot. 

Finally, the interaction of outflow B with the BIMA~4 core leads us to
speculate that outflow B could be triggering the collapse in this core, as has
been proposed in other regions (\eg\ Yokogawa \et\ 2003). This would draw a
scenario in which a YSO in the north of the main cloud is 'responsible' for
star formation in the south, but should be further investigated  by studying in
more detail the morphology and kinematics in the core.


\subsection{YSOs in the region \label{sdsources}}

\subsubsection{Selection of 2MASS sources associated with the region
\label{sd2mass}}

\begin{figure}
\begin{center}
\includegraphics[scale=0.5,angle=0]{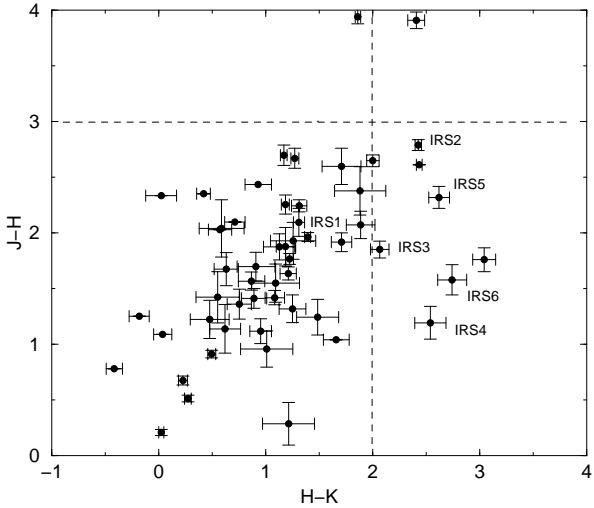}
\caption{
$(J-H)$, $(H-K)$ diagram for the 2MASS sources in the IRAS~20293+3952 region
lying within the BIMA primary beam. We selected those sources, IRS~2 to IRS~6,
with $(H-K)>2$, $(J-H)<3$, and spatially coincident with the dense gas
emission. 
\label{fcolorcolor}
}
\end{center}
\end{figure}

In order to study the different sources associated with the dense gas found
around the \uchii\ region, we extracted a sample of stars within the BIMA
primary beam from the 2MASS Point Source Catalogue (PSC) (Skrutskie \et \
2006), and plotted a $(J-H)$, $(H-K)$ diagram (Fig.~\ref{fcolorcolor}). The
total amount of infrared sources inside the BIMA primary beam is 43. The
color-color diagram shows that there is a bulk of infrared stars at
low-to-moderate values of the color indices, which  occupy the position of
stars with no infrared excess, and a group of stars with high $(H-K)$ and low
$(J-H)$ colors. We took the criteria of $(H-K)>2$, $(J-H)<3$ and spatial
coincidence with emission detected in this work to select those 2MASS sources
that are possibly (but not necessarily) associated with the dense gas around
the \uchii\ region. The selected 2MASS sources are labeled as IRS~1, IRS~2
(following the nomenclature of Kumar \et\ 2002), and IRS~3 to IRS~6
(increasing RA). These infrared sources (with the exception of those that seem
to be background/foreground objects, see below) are listed in
Table~\ref{tsources} together with the compact millimeter sources from Beuther
\et\ (2004b), and the sources found in this work. For each source we show the
main properties derived in this work. 


\subsubsection{Individual YSOs}

\paragraph{IRS~1 and the \uchii\ region:}   From the 2.12 $\mu$m image provided
by Kumar \et\ (2002), we find that IRS~1 is probably a binary system separated
$\sim$3$''$, the fainter component being associated with the \uchii\ region. The
\uchii\ region has a flux density of 7.6~m\jpb\ at 3.6 cm (Sridharan \et \ 2002;
Beuther \et\ 2004a), and a deconvolved size of $4\farcs5$ (0.04~pc). This
corrresponds to an ionizing flux of  $\sim$1.5$\times 10^{45}$ s$^{-1}$,
typical of stars of spectral type B1 (Panagia 1973). Toward the \uchii\ region,
we found the highest rotational temperature in the main cloud, and very low
column densities, as expected  for gas after the passage of an ionization front
(see, \eg\ Dyson \& Williams 1997). 

\begin{table*}{\footnotesize
\caption{List of sources in the I20293 region \label{tsources}}
\centering
\begin{tabular}{lcccccccccc}
\hline\hline
&\multicolumn{2}{c}{Position$^\mathrm{a}$}
&&
&$\Delta v$(\nh)~$^\mathrm{c}$
&$N$(\nh)
&&$T_\mathrm{rot}$
&$I_\nu^\mathrm{peak}$~$^\mathrm{e}$
&$M_\mathrm{c}$~$^\mathrm{f}$
\\
\cline{2-3}
&$\alpha (\rm J2000)$
&$\delta (\rm J2000)$
&&Outf~$^\mathrm{b}$
&(\kms)
&($10^{15}$ cm$^{-2}$)
&Ratio~$^\mathrm{d}$
&(K)
&(m\jpb)
&(\mo)
\\
\hline
\emph{YSOs}\\
IRS~1		     &20:31:11.26     &+40:03:07.6 && NN & -		  & -	 & -   & -	  & 3.9    &0.6       \\
\uchii\                &20:31:11.22     &+40:03:11.0 && NN & $2.4\pm0.3$    & 0.14 & 57  & $38\pm15$& 4.0    &0.4       \\
IRS~2		     &20:31:10.32     &+40:03:16.5 && NN & -		  &	 & -   & -	  & 3.5    &0.5       \\
IRS~3		     &20:31:12.48     &+40:03:20.0 && YN & -		  & -	 & -   & -	  & 9.5    &1.4       \\
IRS~5		     &20:31:13.41     &+40:03:13.7 && ?N & $1.53\pm0.06$  & 0.59 & 42  & $24\pm3$ & 5.2    &1.2       \\
mm1                  &20:31:12.9      &+40:03:22.9 && YY & $1.37\pm0.12$    & 1.1  & 37  & $24\pm3$ & 28.0$^\mathrm{g}$   &4.0       \\
mm2                  &20:31:12.01     &+40:03:12.2 && YN & $1.10\pm0.07$   & 0.88 & 46  & $23\pm4$ & 8.6$^\mathrm{g}$    &1.3       \\
mm3                  &20:31:12.41     &+40:03:10.5 && YN & $1.23\pm0.13$    & 0.66 & 57  & $19\pm1$ & 3.6$^\mathrm{g}$    &0.8       \\
NWS$^\mathrm{h}$     &20:31:13.23     &+40:03:19.5 && NY & $2.00\pm0.11$    & 1.8  & 56  & $29\pm6$ & 3.7    &0.7       \\
SWS$^\mathrm{h}$     &20:31:13.26     &+40:02:58.9 && YN & $\lesssim$1.03$\pm0.04$  & 1.3  & 150 & $20\pm2$ &$<0.28$ &$<0.9$    \\
\emph{Starless Cores}\\
BIMA~3               &20:31:13.92     &+40:03:00.9 && NN & $1.18\pm0.06$  & 5.4  & 290 & $14\pm2$ & 4.0    &2.2       \\
BIMA~4		     &20:31:14.40     &+40:03:07.4 && YY & $1.34\pm0.06$  & 3.8  & 180 & $14\pm2$ & 3.5    &2.0       \\
WCloud$^\mathrm{h}$  &20:31:10.71     &+40:03:30.3 && NN & $\lesssim$1.03$\pm0.02$  & 2.8  & 280 & $15\pm3$ &$< 0.28$ &$< 1.3$    \\
\hline
\end{tabular}
\begin{list}{}{}
\item[$^\mathrm{a}$] Positions correspond to those in the 2MASS PSC for IRS~1 to IRS~5; the centimeter source given by Beuther \et\
(2002b) for the \uchii\ region; the positions given by Beuther \et\ (2004b) for mm1 to mm3; the peak in the rotational temperature map of
Fig.\ref{fpxpx}a for the NWS and the SWS, and the position of the peak in the \nh\ column density map of  Fig.\ref{fpxpx}b for the
starless cores.
\item[$^\mathrm{b}$] High-velocity outflow associated (Y=Yes; N=No): the first character refers to CO from Beuther \et\ (2004a), the second
character refers to high-velocity \chtoh\ from this work.
\item[$^\mathrm{c}$] Line width derived from the fits to the hyperfine structure of \nh(1,1) (see \S\,\ref{satrot}).
\item[$^\mathrm{d}$] Ratio of $N$(\nh) over $N$(\nth).
\item[$^\mathrm{e}$] Intensities at 3.15~mm from this work. The rms noise is 0.7~m\jpb, and the
upper limits for non-detections have been set to 4$\sigma$. 
\item[$^\mathrm{f}$] Associated mass (essentially circumstellar) from the mm continuum emission listed in column (9), assuming a dust emissivity
index $\beta=1$, and estimating the kinetic temperature from the rotational temperature listed in column (8),
following the expression of Tafalla \et\ (2004, see \S\,\ref{sdsources}). 
\item[$^\mathrm{g}$] For mm1, mm2, and mm3 the intensity corresponds to the flux density (in mJy) given by Beuther
\et\ (2004b).
\item[$^\mathrm{h}$] NWS: Northern Warm Spot; SWS: Southern Warm Spot; WCloud:
western cloud (see Fig.~\ref{fm0}b).
\end{list}
}
\end{table*}


\paragraph{IRS~2:} About $10''$ to the northwest of the \uchii\ region, there
is the second brightest infrared source in the field, IRS~2, which shows much
more infrared excess than IRS~1, as already stated by Kumar \et \  (2002). We
could not determine a temperature and column density from \nh \ toward IRS~2
due to the low S/N of the \nh\ emission, but we detected continuum emission at
3.15~mm, and we estimated an associated mass of 0.6~\mo. The high
near-infrared flux of IRS~2 suggests that this is an intermediate-mass YSO.

\paragraph{IRS~3:} IRS~3 is the only infrared source with infrared excess
falling inside the 2$\sigma$ contour level of BIMA~1. The high angular
resolution observations of Beuther \et\ (2004b) reveal a weak extension of mm1
toward the southwest reaching the position of IRS~3. From our data, the
dense gas emission was too weak to derive a temperature and column density
toward this source. Thus, IRS~3 shows strong infrared emission and weak millimeter
emission, indicating that IRS~3 is likely a YSO which is clearing up 
the material in its surroundings. 


\paragraph{IRS~5:} IRS~5 falls inside the 7$\sigma$ contour level of the dust
ridge (see Fig.~\ref{fcontr}), in its eastern edge. We found heating 
associated with IRS~5 (\S~\ref{satrot} and Table~\ref{ttrot}), indicating that
this source is physically associated with the main cloud.  In addition,  the
\nh\ column density is low (Fig.\ \ref{fpxpx}b), and thus IRS~5 could be a
protostar in the process of clearing up the surrounding material.
Alternatively, the low \nh\ column density could be due to depletion of \nh\
onto dust grains.

\paragraph{mm1:} mm1 is likely the main contribution to the emission of BIMA~1,
since it is the strongest compact millimeter source in the region, and is the
driving source of outflow A, which is highly collimated (Beuther \et\ 2004a)
and elongated in the same direction as BIMA~1 (Fig.\ \ref{fcontr}). In
addition, the infrared emission at 2.12~$\mu$m associated with mm1 is faint and
arises mainly from shocked gas (see Fig. 2 of Kumar \et\ 2002). The rotational
temperature toward mm1 is $24 \pm 3$~K (Table~\ref{tsources}), and the \nh\
column density is rather low, possibly due to depletion of \nh\ onto dust
grains. From the flux density of Beuther \et\ (2004b) at 2.6~mm, we obtain a
mass for mm1  of 4.0~\mo, assuming $\beta=1$ (value estimated by Beuther \et\
2004b), and $\Td=32$~K (estimated from the rotational temperature derived in
this work). All this indicates that mm1, in contrast with IRS~3, is deeply
embedded in a massive and not very hot envelope, suggesting that it is an
intermediate/high-mass protostar. 



\paragraph{mm2 and mm3:} Beuther \et\ (2004a) detect two compact millimeter
sources inside the 5$\sigma$ contour level of BIMA~2, mm2 and mm3, and they
suggest that each one is driving a molecular high-velocity CO\,(2--1) outflow
(outflows C and D), which indicates that mm2 and mm3 are truly protostars, and
not heated clumps of dust without any star yet.  Taking the flux density from
Beuther \et\ (2004b) at 2.6 mm, a dust emissivity index $\beta=1$, and
$\Td=30$~K and $\Td=23$~K for mm2 and mm3 respectively ($\Td$ derived from this
work, as in the case of mm1), we estimate associated masses of 1.3~\mo\ for mm2
and 0.8~\mo\ for mm3.


\paragraph{The Northern Warm Spot:}  The Northern Warm Spot is located about
$5''$ (10000~AU) to the southeast of mm1, and is near a local maximum in  \nh\
and \nth\ column densities (see Fig.\ \ref{fpxpx}b and c). From the millimeter
continuum emission we estimated an associated mass of $\sim$0.7~\mo\  (assuming
$\beta=1$ and $\Td=44$~K). Very close to the warm spot, $1\farcs3$ to the east,
there is faint infrared continuum emission at 2.12 $\mu$m, which is not
detected in the $J$ and $H$ band images of 2MASS, indicating that the source
has strong infrared excess. Since there is no H$_2$ line emission at
2.12~$\mu$m tracing shocked gas  (see Fig.~2 of Kumar \et\ 2002), the infrared
emission is most likely tracing an embedded protostar, and not a condensation
heated up by the impact of outflow B. These properties support that the
Northern Warm Spot is the driving source of outflow B.


\paragraph{The Southern Warm Spot:} Similar to the NWS, the Southern Warm Spot
is found $\sim$10$''$ to the southwest of BIMA~3  (see \S \ref{satrot}). At
this position, the temperature enhancement is very significant (see
Table~\ref{ttrot}), and we also found faint infrared emission associated, 
present only in the $K$ filter of 2MASS, and with no H$_2$ line emission
associated. The emission from the continuum at 3.15 mm is below 4$\sigma$,
implying an associated mass $\lesssim 0.9$~\mo\ (assuming $\Td=25$~K, estimated
as in the case of mm1).  Furthermore, there is high-velocity blueshifted
CO\,(2--1) emission spatially coincident with the Southern Warm Spot (Beuther
\et\ 2004a). All this indicates that possibly the heating in the Southern Warm
Spot is due to an embedded low-mass YSO. 



\paragraph{Background/foreground objects:} In addition to IRS~5, we find two
2MASS sources with infrared excess, IRS~4 and IRS~6, spatially coincident with
the dust ridge.   IRS~6 is the source with the highest infrared excess among
the 2MASS sources in the main cloud, suggesting that it is embedded in gas and
dust.   The faint source IRS~4 is located only $4''$ to the southwest of IRS~6,
and also shows infrared excess. However, the association of IRS~4 and IRS~6
with the dense gas is not clear from the rotational temperature and \nh\ column
density maps (Fig.~\ref{fpxpx}a,b), and thus these two infrared sources could
be highly extincted background stars or foreground YSOs belonging to the
cluster.


\paragraph{Summary:}  As seen above, there is a variety of YSOs in the region.
The most massive sources seem to be the \uchii\ region, IRS~1, and IRS~2. To the
east of the massive sources, there is the main cloud of dense gas, where we
find seven sources (IRS~3, IRS~5, mm1, mm2, mm3, the Northern Warm Spot and the
Southern Warm Spot), which seem to be (proto)stars coming from the same natal
cloud. In addition, we found three cores with characteristics similar to
starless cores. The sources seem to be in different evolutionary stages, with
the infrared sources in the more advanced phases, the millimeter sources with
no infrared emission likely being in an earlier stage, and the youngest sources
having the properties of starless cores. We do not find different evolutionary
stages only between the low-mass sources and the high-mass sources (which are
expected because the high-mass sources evolve more rapidly to the main sequence
than the low-mass sources), but among the low-mass sources as well, for which
we expect similar rates of evolution to the main sequence.  For example, IRS~3
and IRS~5 are YSOs strongly emitting in the infrared but with faint millimeter
emission, and do not seem to be driving outflows, suggestive of being at the
end of the accretion phase. Contrary to this, mm2 and mm3 are YSOs with the
millimeter emission much stronger than the infrared emission, and they are the
driving sources of collimated outflows, indicative of being in the main
accretion phase. The masses of IRS~3 and IRS~5 are difficult to estimate, but
comparing with the infrared emission of the known intermediate/high-mass
sources one may classify them as low-mass YSOs. The masses of mm2 and mm3, 
estimated from the millimeter emission, are around 1~\mo. Thus, these low-mass
YSOs (IRS~3/IRS~5, and mm2/mm3) seem to be in different evolutionary stages,
and we conclude that stars are not forming simultaneously in this cluster
environment. Rather, there may be different generations, as found toward other
star-forming regions (\eg\ NGC\,6334: Beuther \et\ 2005b; S235A-B: Felli \et\
2006; L1551: Moriarty-Schieven \et\ 2006).


\subsection{Spatial distribution of the YSOs in the region \label{sdspatial}}

In this section we consider whether interaction between the different sources
is important in the determination of the spatial distribution of the YSOs. We
find that star formation is localized in the north of the main cloud, where
there are around six YSOs, while in the south we find sources with properties
of starless cores, and only one YSO.

We consider first whether the \uchii\ is responsible for such a spatial
distribution. In section \ref{sdinter} we discussed some evidence of
interaction of the \uchii\ region with the edges of the main cloud, mainly with
BIMA~2. This suggests that mm2 and mm3 could have been triggered by the \uchii\
region. If this was the case, one would expect to find some evidence of a
compression front expanding away from the \uchii\ region. In the p-v plot of
Fig.~\ref{fpvwestcloud}, the \nth\ emission shows the shape of an incomplete
ring, which could be interpreted as an incomplete expanding shell. However, the
velocity field as seen from \nth\ in Fig.~\ref{fn2hm1m2}b does not show any
radial symmetry with respect to  the \uchii\ region, and the presence of outflow
B makes difficult to disentangle the advance of any compression front. In
addition, the formation of the YSOs in the northeast of the main cloud is not
likely due to triggering by the \uchii\ region, since they are located farther
away than mm2 and mm3 and are not in earlier evolutionary stages. Therefore,
triggering by the \uchii\ region does not seem to be the dominant agent causing
star formation in this clustered environment.

Another possibility for the formation of the YSOs in the north of the main
cloud could be a merging of the main cloud with the western cloud. In
Fig.~\ref{fpvwestcloud}, we found that the western cloud has an extension
intersecting the main cloud, seen at velocities $>7$ \kms\ in the channel maps
(Fig.~\ref{fchn2h}).   Merging of two clouds has been proposed as a mechanism
to trigger star formation in other regions, (Wiseman \& Ho 1996; Girart \et\
1997; Looney \et\ 2006; Peretto \et\ 2006). Note however that no further
evidence of such a merging (heating, line broadening, two clear velocity
components across the main cloud) is seen associated with the extension of the
western cloud, and thus the merging scenario is not fully consistent with our
data.


We therefore conclude that the spatial distribution of the YSOs in this cluster
environment could simply reflect the initial conditions of the cloud: if the
main cloud was originally much denser in the north than in the south, we would
possibly observe a similar situation to what we have found.

\section{Conclusions}

We observed with the BIMA and VLA arrays the continuum emission at 3.15~mm, and
the \nth, \nh, and \chtoh\ emission toward IRAS~20293+3952, a region in which
star formation is taking place in a closely-packed environment. Our main
conclusions can be summarized as follows:

\begin{enumerate} 

\item The dense gas traced by \nth\ and \nh\ shows two different clouds, one to
the east of the \uchii\ region (main cloud), of $\sim$0.5~pc of size and
$\sim$250~\mo, and another cloud to the northwest (western cloud), of
$\sim$0.15~pc and $\sim$30~\mo, and redshifted with respect to the main cloud.
The dust emission reveals two strong components in the northern side of the
main cloud,  BIMA~1 and BIMA~2, and two fainter components in the southern
side, BIMA~3 and BIMA~4, together with extended dust emission forming a common
envelope. Regarding the \chtoh, we found strong emission in a fork-like
structure associated with outflow B from Beuther \et\ (2004a), as well as two
\chtoh\ clumps associated with outflow A.

\item We found that the rotational temperature is higher in the northern side
of the main cloud, around 22~K, than in the southern side, around 16 K. In
contrast, the \nh\ column density has the highest values in the south of the
main cloud. The \nth\ column density distribution resembles the dust emission,
strong in the northern side of the main cloud. We found three local temperature
enhancements which seem to be associated with embedded YSOs, one of them
associated with a 2MASS source, and the other two, the Northern Warm Spot and
the Southern Warm Spot, associated with faint continuum infrared emission at
2.12~$\mu$m.

\item There is strong chemical differentiation in the region. In particular, we
found low values of the \nh/\nth\ ratio, $\sim$50, associated with YSOs, and
high values, up to 300, associated with starless cores. This is consistent with
\nh\ being enhanced with respect to \nth\ at moderate densities ($\lesssim
10^5$~\cmt), while at densities around $\sim$10$^6$~\cmt\ the \nh\ enhancement
may be less important (due to shorther lifetimes of the cores), and
additionally the effects of \nh\ depletion may start to play a role.

\item We identified three cores in the \nh\ column density map, with low
temperatures and with no infrared emission associated, similar to  starless
cores in low-mass star-forming regions, but with temperatures slightly higher,
$\sim$15~K, possibly due to external heating from the nearby OB stars. These
are associated with BIMA~3, BIMA~4, and the western cloud.

\item Interaction between the different sources in the region is important.
First, the \uchii\ region is interacting with the main cloud heating and
enhancing the CN\,(1--0) emission in the edge of the main cloud that is facing
the \uchii\ region. Second, one of the outflows in the region, outflow A, seems
to be excavating a cavity and heating its walls. Third, another outflow in the
region, outflow B, is interacting with the BIMA~4 starless core, likely
producing the deflection of the outflow seen at this position. Such a shock
could be illuminating a clump with narrow \chtoh\ lines located $\sim$0.2~pc to
the northeast of the shock. We propose that the Northern Warm Spot may contain
the driving source of outflow~B. The properties of the molecular gas around the
outflows suggest that outflow A is more energetic and/or older than
outflow~B.  

\item There are about eight YSOs in the dense gas near the \uchii\ region. For
the YSOs for which we could estimate the associated circumstellar mass, we
found values ranging from $<0.4$ to $\sim$4~\mo. In addition, the YSOs seem to
be in different evolutionary stages, even if we consider only the low-mass
sources. Thus, stars do not seem to form simultaneously in this cluster
environment.

\item While we cannot discard that interaction between the different sources
may have triggered star formation in particular cases, triggering cannot
explain the overall spatial distribution of the YSOs, suggesting that this may
be essentially determined by the initial conditions in the cloud.


\end{enumerate}

\begin{acknowledgements}

A.\ P. is grateful to M.\ S. Nanda Kumar for providing us the H$_2$ images, to
Maite Beltran and Rosario L\'opez for help in technical aspects of this paper,
to Serena Viti for chemistry modeling, and to Oscar Morata  and the
anonymous referee, whose comments and discussions significantly improved the
clarity of the paper. The authors  A.\ P., R.\ E., and J.\ M.\ G. are
supported by a MEC grant AYA2005-08523 and FEDER funds. H.\ B. acknowledges
financial support by the Emmy-Noether-Program of the Dutsche
Forschungsgemeinschaft (DFG, grant BE2578). This publication makes use of data
products from the Two Micron All Sky Survey, which is a joint project of the
University of Massachusetts and the Infrared Processing and Analysis
Center/California Institute of Technology, funded by the National Aeronautics
and Space Administration (NASA) and the National Science Foundation. 


\end{acknowledgements}


\end{document}